\documentclass[%
 reprint,
superscriptaddress,
 amsmath,amssymb,
 aps,
pra,
]{revtex4-2}
\pdfoutput=1

\usepackage{graphicx}
\usepackage{dcolumn}
\usepackage{bm}
\usepackage{natbib}
\usepackage{textcase}
\usepackage{url}
\usepackage{amsmath}

\begin{document}

\preprint{APS/123-QED}

\title{Covariance Distributions in Single Particle {\rm Tr}acking}

\author{Mary Lou P Bailey}
\affiliation{%
	Integrated Graduate Program in Physical and Engineering Biology, Yale University, New Haven, Connecticut 06520, USA
}
\affiliation{%
	Department of Applied Physics, Yale University, New Haven, Connecticut 06520, USA
}%
\author{Hao Yan}%
\affiliation{%
	Integrated Graduate Program in Physical and Engineering Biology, Yale University, New Haven, Connecticut 06520, USA
}
\affiliation{%
 Department of Physics, Yale University, New Haven, Connecticut 06520, USA
}%
\author{Ivan Surovtsev}%
\affiliation{%
 Department of Physics, Yale University, New Haven, Connecticut 06520, USA
}%
\affiliation{%
	Department of Cell Biology, Yale School of Medicine, New Haven, Connecticut 06520, USA
}%
\affiliation{%
	Department of Physics, Yale University, New Haven, Connecticut 06520, USA\\
}%
\author{Jessica F Williams}%
\affiliation{%
	Department of Cell Biology, Yale School of Medicine, New Haven, Connecticut 06520, USA
}%
\author{Megan C King}%
\affiliation{%
	Integrated Graduate Program in Physical and Engineering Biology, Yale University, New Haven, Connecticut 06520, USA
}
\affiliation{%
	Department of Cell Biology, Yale School of Medicine, New Haven, Connecticut 06520, USA
}%
\author{Simon G J Mochrie}%
 \email{simon.mochrie@yale.edu}
 \affiliation{%
	Integrated Graduate Program in Physical and Engineering Biology, Yale University, New Haven, Connecticut 06520, USA
}
\affiliation{%
	Department of Applied Physics, Yale University, New Haven, Connecticut 06520, USA
}%
\affiliation{%
 Department of Physics, Yale University, New Haven, Connecticut 06520, USA
}%

%

\date{\today}

\begin{abstract}
Several recent experiments, including our own experiments in the fission yeast, {\em S. pombe}, have characterized the motions of gene loci within living nuclei by measuring the locus position over time, then proceeding to obtain the statistical properties of this motion. To address the question of whether a population of such single particle tracks, obtained from many different cells, corresponds to a single mode of diffusion, we derive theoretical equations describing the probability distribution of the displacement covariance, assuming the displacement itself is a zero-mean multivariate Gaussian random variable. We also determine the corresponding theoretical means, variances, and third central moments. Bolstering the theory is good agreement between its predictions and the results obtained for various simulated and measured data sets, including simulated particle trajectories undergoing simple  and anomalous diffusion, 
and the measured trajectories of an optically-trapped bead in water, and in a viscoelastic polymer solution. We also show that, for sufficiently long tracks, each covariance distribution in all of these examples is well-described by a skew-normal distribution with mean, variance, and skewness given by the theory. However, for the experimentally measured motion of a gene locus in {\em S. pombe}, we find that the first two covariance distributions are wider than predicted, although the third and subsequent covariance distributions are well-described by theory. This observation suggests that the origin of the theory-experiment discrepancy in this case is associated with localization noise, which influences only the first two covariances. Thus, we hypothesized that the discrepancy is caused by locus-to-locus heterogeneity in the localization noise, of independent measurements of the same tagged site. Indeed, simulations implementing heterogeneous localization noise revealed that the excess covariance widths can be largely recreated on the basis of heterogeneous noise. Thus, we conclude that the motion of gene loci in fission yeast is consistent with a single mode of diffusion.

\end{abstract}

\maketitle


\section{\label{sec:level1}Introduction}

Single particle tracking has long been been applied to elucidate the dynamics of various soft-matter and biological systems
\cite{Crocker1996,Crocker2000,Dahan2003,Weeks2000,Parry2014}.
Recent advances in fluorescent tagging and imaging now also enable tracking-based studies of single molecules
or moieties within living cells.
The motion of fluorescently labeled particles is most often analyzed by determining the mean squared displacement (MSD) as a function of time delay between observations. This approach has been applied to a wide variety of macro- and supra-molecular complexes inside cells from diverse organisms \cite{Heun2001,Vrljic2002,Manley2008,Hajjoul2013,Normanno2014,Koo2015,Hosy2015,Ye2016,Liu2018,Martens2019,Vink2020,Rey-Suarez2020}.
Examples in mammalian cells include reports of  simple diffusion for a transmembrane protein \cite{Vrljic2002} as well as multiple diffusive states within subpopulations (immobilized, subdiffusive and diffusive) for a viral protein \cite{Manley2008} or several sub-populations (immobilized, slow and fast diffusion) for a DNA-binding protein \cite{Normanno2014}.
For the DNA itself, subdiffusion has been reported for DNA loci in bacteria \cite{Weber2010} and subdiffusion \cite{Hajjoul2013} or confined  diffusion \cite{Marshall1997,Heun2001} in eukaryotes. 

The question of how to optimally determine the diffusivity from the time-averaged MSD (taMSD) has been investigated thoroughly for the case of simple diffusion \cite{Martens2019,Vink2020,Qian1991,Saxton1997,Michalet2010,Michalet2012}.
Typically, the slope of the taMSD for a limited number of time delays is used to yield an estimate of the diffusivity for each track and distribution of diffusivities across multiple tracks.
However, within a track, the displacements measured for different time delays are not statistically independent.
This, together with the particle position measurements being imperfect, as they include static localization noise (the uncertainty in particle position due to a limited number of detected photons) \cite{Thompson2002,Savin2005,Mortensen2010} and motion blur (the spatial spread of the signal due to camera integrating a particle's positions over exposure time) \cite{Savin2005}, confounds MSD analysis.
Recently, Vestergaard et al. \cite{Vestergaard2014} have shown that the optimal way to gauge diffusivity, while accounting for localization noise and motion blur, is to use an estimator based on the covariances of the particle displacements.

Another fundamental question is whether a single or multiple diffusion coefficients are appropriate to describe data collected from heterogeneous biological systems.
Several approaches have been developed to sort particle trajectories into different diffusive states \cite{Persson2013,Koo2015,Koo2016,Martens2019,Rey-Suarez2020}. 
However, the simplest way to investigate whether data realize a single diffusivity or not is to compare the width of the measured diffusivity distribution to the width one would expect for a single diffusivity. 

Vestergaard et al.\cite{Vestergaard2014} derived the expected distribution and variance of diffusivities measured from tracks of finite length in the case of simple diffusion, assuming that the static localization error is Gaussian-distributed.
However, this analysis is not applicable to the more general case of anomalous diffusion.
In the case of simple diffusion analyzed in Ref.~\cite{Vestergaard2014}, there are only two non-zero covariances, linearly related to the diffusivity. By contrast,  there are as many non-zero covariances as there are steps in a track in the case of anomalous diffusion.
A biologically important case of anomalous diffusion is the motion of a chromosomal locus. 
While the mean covariances for anomalous diffusion have been analyzed \cite{Weber2012,Backlund2015}, to-date we are unaware of any comparison between the measured and predicted covariance distributions in the context of gene loci motion, nor any consideration of whether the motion of gene loci is homogeneous in time or whether a gene locus may undergo transitions among different modes of diffusion. 
The goal of this paper is to answer the question: Does the  {\em in vivo} motion of a gene locus in
fission yeast follow a single mode of diffusion, or not? 

Specifically, this paper focuses on the expected scatter in SPT track descriptors for a single diffusive state in order to be able to identify additional scatter that may arise as a result of variations in the underlying dynamics. To this end, we calculate the probability distribution function of these elements, expressed as a Fourier transform, which we perform numerically for comparison to data and simulations. However, we find that the exact probability density can be well approximated by the skew normal distribution. It follows that fitting measured covariance distributions to the skew normal provides the important descriptors of the distribution, namely the mean, variance, and third central moment, which can then be compared to the exact theoretical values of these quantities. Agreement between the two strongly suggests that the particle tracks in question exhibit one mode of diffusion. 

The paper is organized as follows. In Sec.~\ref{1DSection} and Sec.~\ref{2DSection}, we present the probability distribution, the mean, the variance, and the third central moment of the displacement covariance matrix elements for arbitrary modes of diffusion in one and two dimensions (1D and 2D), respectively.
 In Sec.~\ref{2Ddiffusion}, we specialize to give the diffusivity distribution, its variance, and third central moment in the case of simple 2D diffusion, reproducing some of the results of Ref.~\cite{Vestergaard2014} via a different route.
In Sec.~\ref{Sec:Comparisons}, we compare  theoretical
covariance distributions for a zero-mean Gaussian random process, expressed in terms of the mean covariance matrix elements,  to a number of simulated and experimentally measured covariances. For simulated simple diffusion (Sec.~\ref{sec:simpleD}) we find excellent agreement with theory.  We also apply our methods to optical tweezers data (Sec.~\ref{sec:Beads}), and again find good agreement with theory. 
The agreement with theory in this case  underscores that the usual procedures for fitting optical tweezers measurements should be generally modified to account for the motion blur, as pointed out previously \cite{Wong2006}.
Finally, we analyze motion of gene loci in living {\em S. pombe} yeast cells (Sec.~\ref{sec:bio}), and find an important discrepancy with theory. Our analysis points to the discrepancy stemming from experimental locus-to-locus heterogeneity in localization noise,
while the theory assumes the localization error to be drawn from a single normal distribution. After accounting for this heterogeneity with simulations, through implementing heterogeneous localization noise,  we conclude that the gene loci in living {\em S. pombe} indeed undergo a single mode of diffusion. Finally, in Sec. IV, we summarize
and conclude.


\section{Theoretical distribution of covariance matrix elements}
\label{Sec:Theory}
\subsection{One-dimensional analysis}
\label{1DSection}


We consider data that consist of a collection of single particle tracks,
each characterized by the same well-defined diffusive properties.
There are at least three important reasons to then consider tracks of finite length.
The first is that over time, fluorescently-labeled proteins {\em in vivo} may experience changes in their diffusive properties -- changes in their diffusive state --   such as those caused by binding and unbinding events, for example, to DNA or other complexes in the cell.
The second is that, even for proteins that always remain in a single diffusive state, individual fluorescent labels can blink and eventually  bleach.
Third, in microscopy experiments, proteins can readily diffuse out of the  focal volume. 
All of these processes give rise to tracks that realize a single diffusive state for a finite duration.
Therefore, we start off by considering a population of tracks
each comprising $N+1$ particle coordinates along the $x$-axis, $\{ x_j\}$, with $j=1$ through $N$,
corresponding to $N$ particle displacements along the $x$-axis, $\{ \Delta x_j = x_{j+1}-x_{j} \}$,
and we focus on the covariance of these one-dimensional displacements.
Each measured track provides an
estimate of the covariance matrix elements, which, assuming that the diffusive properties do not vary within a
track, should depend only on time separation $n$ for a stationary process.
Each measured track provides an estimate of the covariance matrix elements
via:
\begin{equation}
S_n=\frac{1}{N-n} \sum_{j=1}^{N-n} \Delta x_j \Delta x_{j+n},
\label{ExpCov}
\end{equation}
where $\Delta x_j$ is the displacement of step $j$ and
$N$ is the total number of steps in the track.
Because of their inherent stochasticity,
different tracks yield different values for  $S_n$.
However, averaging over many tracks
yields the underlying mean covariance matrix:
\begin{align}
\mathbf{\Sigma}
& =& \left( \begin{array}{ccccc}
\left < S_0 \right >& \left <S_1  \right >& \left <S_2  \right >&\left < S_3  \right >& ....\\
\left <S_1  \right >& \left <S_0  \right >& \left <S_1  \right >&\left < S_2  \right >& ....\\
\left <S_2 \right > & \left < S_1  \right >&\left < S_0  \right >& \left <S_1  \right >& ....\\
 \left <S_3  \right >& \left <S_2 \right > & \left <S_1  \right >& \left <S_0  \right >& ....\\
. & . & . & . & .... \\
. & . & . & . & .... \\
. & . & . & . & .... \\
 \end{array} \right) \nonumber \\
& =&
 \left( \begin{array}{ccccc}
\Sigma_0 & \Sigma_1 & \Sigma_2 & \Sigma_3& ....\\
\Sigma_1 & \Sigma_0 & \Sigma_1 & \Sigma_2& ....\\
\Sigma_2 & \Sigma_1 & \Sigma_0 & \Sigma_1& ....\\
\Sigma_3 & \Sigma_2 & \Sigma_1 & \Sigma_0& ....\\
. & . & . & . & .... \\
. & . & . & . & .... \\
. & . & . & . & .... \\
 \end{array} \right ),
\label{Toeplitz}
\end{align}
where the angular brackets indicate an ensemble average over
tracks, and we have defined $\Sigma_n = \left < S_n \right >$.
In general, the covariance matrix is a symmetric Toeplitz matrix. We note here that through this paper, "mean" is used in two different senses. In the context of SPT, "mean" refers to averaging over the ensemble of the tracks, while in the context of the theoretical probability distribution function, "mean" refers to the calculated expected values.

Eq.~\ref{ExpCov} permits us to calculate $S_n$ for each individual track, and thus, to determine the experimental distribution $S_n$ from a population of experimental or simulated tracks.
We can then test  our understanding of the underlying
diffusive process
 by comparing these empirical distributions to corresponding
theoretical predictions.

To determine theoretical expressions for the distributions of $S_n$, 
our starting point is
the hypothesis that the probability
of observing a particular $N$-step track is given by a multivariate Gaussian distribution,
characterized by the $N \times N$
covariance matrix $\mathbf{\Sigma}$ (Eq.~\ref{Toeplitz}):
\begin{align}
P(\Delta \mathbf{x}|\mathbf{\Sigma}) &=& \frac{1}{( 2 \pi)^{N/2} | \mathbf{\Sigma}|^{1/2}}  \exp{\left[-\frac{1}{2}  \Delta \mathbf{x}^T \mathbf{\Sigma}^{-1} \Delta \mathbf{x} \right]},
\label{multivariate}
 \end{align}
where 
$\Delta \mathbf{x} = (\Delta x_1, \Delta x_2, ... \Delta {x}_N)^T$ is the vector of $N$ successive particle displacements along the $x$-direction within the track,
and
$|\mathbf{\Sigma}|$ is the determinant of the covariance matrix.
To proceed, we introduce
the (Toeplitz) matrices,
$[{\bf C}_0]_{j k} = \frac{2}{N} {\bf I}$
and
$[{\bf C}_n]_{j k} = \frac{1}{(N-n)} \delta_{j~ k\pm n}$,
which permits us to
re-write Eq.~\ref{ExpCov} as
\begin{equation}
S_0=\frac{1}{N} \sum_{j=1}^{N} \Delta x_j \Delta x_{j}
=\frac{1}{2}  \Delta {\bf x}^T {\bf C}_0 \Delta {\bf x},
\label{EQC0}
\end{equation}
and
\begin{equation}
S_n=\frac{1}{N-n} \sum_{j=1}^{N-n} \Delta x_j \Delta x_{j+n}
=\frac{1}{2}  \Delta {\bf x}^T {\bf C}_n \Delta {\bf x},
\label{EQC1}
\end{equation}
for $n>0$.

Using Eq.~\ref{multivariate}, Eq.~\ref{EQC0}, and Eq.~\ref{EQC1},
and the Fourier transform representation of the Dirac delta function,
we may express the probability distribution of
the random variables,
$S_n$, for a given covariance matrix, in terms of a certain matrix determinant:
\begin{widetext}
\begin{equation}
	\begin{array}{ll}
P(S_n | \mathbf{\Sigma} ) &= \int d(\Delta x_1) d(\Delta x_2) ... d(\Delta x_N)  P(\Delta {\bf x} | \mathbf{\Sigma} )
\delta \left (S_n - \frac{1}{2}  \Delta {\bf x}^T {\bf C}_n \Delta {\bf x} \right ) \\[10pt]

&= \int^\infty_{-\infty} \frac{d \omega}{2 \pi } \int d(\Delta x_1) d(\Delta x_2) ... d(\Delta x_N)
\frac{1}{\sqrt{(2 \pi)^{N} |\mathbf{\Sigma}|}}
e^{i \omega S_n -
\frac{1}{2}
 \Delta {\bf x}^T 
 \left (\mathbf{\Sigma}^{-1}+  i \omega {\bf C}_n \right )  \Delta {\bf x}  }\\[10pt]
&= 
  \int^\infty_{-\infty} \frac{d \omega}{2 \pi }
\frac{1}{  \sqrt{ |\mathbf{\Sigma}| | \mathbf{\Sigma}^{-1}+ i \omega {\bf C}_n | }}
e^{i \omega S_n}\\[10pt]
&= 
  \int^\infty_{-\infty} \frac{d \omega}{2 \pi }
\frac{1}{  \sqrt{ |{\bf I}+  i \omega \mathbf{\Sigma} {\bf C}_n | }}
e^{i \omega S_n},
\end{array}
\label{1DProbabilityDensity}
\end{equation}
\end{widetext}
where
$| \mathbf{\Sigma}^{-1}+ i \omega {\bf C}_n |$
and $| \mathbf{I}+ i \omega  \mathbf{\Sigma}  {\bf C}_n|$ are the determinants of
$\mathbf{\Sigma}^{-1}+ i \omega {\bf C}_n $ and $\mathbf{I}+ i \omega  \mathbf{\Sigma} {\bf C}_n$, respectively.
Eq.~\ref{1DProbabilityDensity} is the probability density of $S_n$, given the covariance matrix ${\mathbf \Sigma}$.
After calculating the determinant as a function of $\omega$,
the integral over $\omega$ may be performed numerically
to obtain  $p(S_n | {\mathbf \Sigma})$.
%

In Appendix \ref{ndspace}, we show from Eq.~\ref{1DProbabilityDensity} that
 the first three moments of $S_n$ are
\begin{equation}
\left < S_n \right > =
\frac{1}{2}{{\rm Tr}( {\mathbf \Sigma} {\bf C}_n)},
\label{EQ7}
\end{equation}
\begin{equation}
<S_n^2> =
\frac{({\rm Tr}{\mathbf \Sigma} {\bf C}_n)^2}{4}  +\frac{{\rm Tr}({\mathbf \Sigma} {\bf C}_n)^2}{2},
\label{EQ8}
\end{equation}
and
\begin{equation}
\left < S_n^3 \right > = \frac{1}{8} ({\rm Tr}{\mathbf \Sigma} {\bf C}_n)^3 +\frac{3}{4} {\rm Tr}({\mathbf \Sigma} {\bf C}_n){\rm Tr}({\mathbf \Sigma} {\bf C}_n)^2 +{\rm Tr}({\mathbf \Sigma} {\bf C}_n)^3.
\label{EQ9}
\end{equation}
%
%
%
%
%
%
Combining Eq.~\ref{EQ7} and Eq.~\ref{EQ8}, we find that the variance of $S_n$ is
\begin{equation}
\sigma^2_{S_n} = \left < S_n^2 \right > - \left < S_n \right > ^2
=
\frac{{\rm Tr}({\mathbf \Sigma} {\bf C}_n)^2}{2}.
\label{Variance1D}
\end{equation}
Combining Eq.~\ref{EQ7}, Eq.~\ref{EQ8}, and Eq.~\ref{EQ9}, we find that the third central moment of $S_n$ is
\begin{equation}
\mu_3 = \left < (S_n - \left < S_n \right >) ^3 \right > =
{\rm Tr}({\mathbf \Sigma} {\bf C}_n)^3.
\label{Skewness1D}
\end{equation}


\subsection{Two-dimensional analysis}
\label{2DSection}
Single-particle tracking often results in two-dimensional data, so that two vectors are available for each track, namely $\Delta{\bf x}$
and  $\Delta{\bf y}$.  In this case, given that the motion along each dimension is independent of the other,  the likelihood of observing
a particular trajectory is
\begin{widetext}
\begin{equation}
P(\Delta \mathbf{x},\Delta \mathbf{y}|\mathbf{\Sigma}_x,\mathbf{\Sigma}_y) = \frac{1}{( 2 \pi)^{N} | \mathbf{\Sigma}|}  \exp{\left[ -\frac{1}{2} \Delta \mathbf{x}^T \mathbf{\Sigma}_x^{-1} \Delta \mathbf{x} -\frac{1}{2} \Delta \mathbf{y}^T \mathbf{\Sigma}_y^{-1} \Delta \mathbf{y} \right]},
\label{multivariate2}
 \end{equation}
 \end{widetext}
 where $\mathbf{\Sigma}_{x}$ and $\mathbf{\Sigma}_{y}$ are the covariance matrices for $x$- and $y$-displacements, respectively.
If we redefine $S_n$  to correspond to two dimensions via
  \begin{equation}
S_n=\frac{1}{N-n} \sum_{j=1}^{N-n} \Delta x_j \Delta x_{j+n} + \frac{1}{N-n} \sum_{j=1}^{N-n} \Delta y_j \Delta y_{j+n},
 \end{equation}
then, correspondingly, we can introduce a two-dimensional covariance matrix, ${\mathbf \Sigma}$, via
 \begin{equation}
 \mathbf{\Sigma}
= \mathbf{\Sigma}_x + \mathbf{\Sigma}_y.
 \end{equation}
Commonly, the diffusive behavior is isotropic, in which case we furthermore have that
 \begin{equation}
 \mathbf{\Sigma}_x =   \mathbf{\Sigma}_y =\frac{1}{2} \mathbf{\Sigma}.
 \end{equation}
 In this isotropic case, it follows that
 the probability density for the two-dimensional covariance is (calculated similarly to Eq.~\ref{1DProbabilityDensity})
 \begin{widetext}
 \begin{equation}
        \begin{array}{ll}
P(S_n | \mathbf{\Sigma} ) &= \int d(\Delta x_1)  ... d(\Delta y_1)...  P(\Delta {\bf x}, \Delta {\bf y}  | \mathbf{\Sigma} )
\delta \left (S_n - \frac{1}{2}  \Delta {\bf x}^T {\bf C}_n \Delta {\bf x} -  \frac{1}{2}  \Delta {\bf y}^T {\bf C}_n \Delta {\bf y} \right ) \\[10pt]
 &= 
  \int^\infty_{-\infty} \frac{d \omega}{2 \pi }
\frac{1}{  { |{\bf I}+  \frac{i}{2} \omega \mathbf{\Sigma} {\bf C}_n |  }}
e^{i \omega S_n}.
\end{array}
\label{EQ-16}
\end{equation}
\end{widetext}


From Appendix \ref{ndspace}, we calculate $\left < S_n \right >$, $\left < S_n^2 \right >$, and $\left < S_n^3 \right >$ in this two-dimensional case.
Using these results, it is straightforward to show that
the mean, variance, and third central moment of the  $S_n$ in 2D are
\begin{equation}
\left < S_n \right > = \frac{1}{2}{{\rm Tr}( {\mathbf \Sigma} {\bf C}_n)},
\label{mean}
\end{equation}
\begin{equation}
\sigma^2_{S_n} = \frac{1}{4}{\rm Tr}({\bf {\mathbf \Sigma} {\bf C}}_n)^2,
\label{Variance2D}
\end{equation}
and
\begin{equation}
\left < (S_n-\left <  S_n \right > )^3 \right >  =\frac{1}{4}{\rm Tr}({\mathbf \Sigma} {\bf C}_n)^3.
\label{Skewness2D}
\end{equation}


\subsection{Simple 2D diffusion}
\label{2Ddiffusion}
Many experimental systems can be expected to realize simple 2D diffusion with experimental errors,
corresponding to a tridiagonal covariance matrix, where the only non-zero covariance matrix elements are $\Sigma_0$ and $\Sigma_1$,
which are related to the diffusion coefficient, $D$,  the static localization noise, $\sigma^2$, the time between camera exposures, $\Delta t$, and the exposure time,
$\Delta t_E$, via \cite{Vestergaard2014}
\begin{equation}
\Sigma_0 = 4D \Delta t -\frac{4}{3}D \Delta t_E +2 \sigma^2
\label{EQ-26}
\end{equation}
and
\begin{equation}
\Sigma_1 = -\sigma^2 +\frac{2}{3} D \Delta t_E.
\label{EQ-27}
\end{equation}
The static localization noise,  $\sigma^2$,  is the error in particle localization that results from counting a limited number of photons.
The terms involving the exposure time, $\Delta t_E$, correspond to motion blur, because the particle position is integrated while the shutter is open.

In this case,
 $\frac{1}{4}{\rm {\rm Tr}} ( \mathbf{\Sigma} \mathbf{C}_n)^2$ and $\frac{1}{4}{\rm {\rm Tr}} ( \mathbf{\Sigma} \mathbf{C}_n)^3$ can be
 evaluated in closed form
with the results that
\begin{equation}
\sigma^2_{S_0}= \frac{\Sigma_0^2+(2-\frac{2}{N})\Sigma_1^2}{N},
\label{EQ-20}
\end{equation}
\begin{equation}
\sigma^2_{S_1}= \frac{\Sigma_0^2+(3-\frac{2}{N-1})\Sigma_1^2}{2(N-1)},
\label{EQ-21}
\end{equation}
\begin{equation}
\sigma^2_{S_n}= \frac{\Sigma_0^2+(2-\frac{2}{N-n})\Sigma_1^2}{(N-n)},
\label{EQ-22}
\end{equation}
for $n>1$,
\begin{equation}
\left < (S_0-\left <S_0 \right>)^3 \right >
= \frac{2N\Sigma_0^3+(12N-12)\Sigma_0\Sigma_1^2}{N^3},
\label{EQ-23}
\end{equation}
\begin{equation}
\left < (S_1-\left <S_1 \right>)^3 \right >
= \frac{(10N-22)\Sigma_1^3+(9N-15)\Sigma_0^2\Sigma_1}{2(N-1)^3},
\label{EQ-24}
\end{equation}
\begin{equation}
\left < (S_2-\left <S_2 \right>)^3 \right > = \frac{((N-30) \Sigma_1^2 \Sigma_0}{2(N-2)},
\label{EQ-25}
\end{equation}
\begin{equation}
\left < (S_3-\left <S_3 \right>)^3 \right > = \frac{(3N-15) \Sigma_1^3}{2(N-3)^2},
\label{EQ-25}
\end{equation}
\begin{equation}
\left < (S_n-\left <S_n \right>)^3 \right > = 0,
\label{EQ-25}
\end{equation}
for $n \geq 4$.
Eqs.~\ref{EQ-20}, \ref{EQ-21}, and \ref{EQ-22} reproduce Eq.~9 of Ref.~\cite{Vestergaard2014}, which employs a different method to find the variance of $S_n$.

Eqs.~\ref{EQ-26} and \ref{EQ-27} can be used to estimate $D$ from each individual track. For a given $N$-displacement track we have: 

\begin{widetext}
	\begin{equation}
	D=\frac{1}{4}\left(\frac{S_0}{\Delta t}+\frac{2 S_1}{\Delta t}\right)=
	\frac{ \Delta {\bf x}^T {\bf C}_0  \Delta {\bf x} }{4 \Delta t} 
	+
	\frac{ \Delta {\bf y}^T {\bf C}_0  \Delta {\bf y} }{4 \Delta t} 
	+ \frac{ \Delta {\bf x}^T {\bf C}_1 \Delta {\bf x} }{2 \Delta t}
	+ \frac{ \Delta {\bf y}^T {\bf C}_1 \Delta {\bf y} }{2 \Delta t}.
	\label{EQ-28}
	\end{equation}

Because of the physical importance of the diffusion coefficient, we also consider its distribution, mean, variance and third moment for a distribution of D values estimated from tracks using Eq.~\ref{EQ-28}.

It follows that the
the probability density, the variance, and the third central moment of the diffusion coefficient are
 \begin{equation}
        \begin{array}{ll}
P(D |  \mathbf{\Sigma} ) = 
  \int^\infty_{-\infty} \frac{d \omega}{2 \pi }
\frac{1}{  { |{\bf I}+  \frac{i}{8 \Delta t} \omega  \mathbf{\Sigma} ({\bf C}_0+2 {\bf C}_1) |  }}
e^{i \omega D},
\end{array}
\label{DistD}
\end{equation}

\begin{align}
\label{EQ-34}
\sigma_D^2&=\frac{1}{4}{\rm Tr}[{\mathbf \Sigma} (\frac{{\bf C}_0+2{\bf C}_1}{4\Delta t})]^2=\frac{ \frac{1}{4} \left(\frac{3N-1}{N(N-1)} \Sigma_0^2+\frac{8}{N}\Sigma_0 \Sigma_1
+\frac{8N^3-16N^2+6N-2}{N^2(N-1)^2} \Sigma_1^2 \right)}{4 (\Delta t)^2} \\ \nonumber
&=
 \frac{
	\frac{D^2}{18}(2N - 1)(1+N(22N-25))
	+\frac{D}{6} \frac{\sigma^2}{\Delta t}(4N^3 - 3N + 1)
	+\frac{1}{8} \left( \frac{\sigma^2}{\Delta t} \right)^2 (2N^3 - 3N - 1)
}
{
	(N-1)^2 N^2
},
\end{align}
and
\begin{align}
\label{EQ-35}
\left < (D-\left < D \right >)^3 \right > &=\frac{1}{4}{\rm Tr}[{\mathbf \Sigma} (\frac{{\bf C}_0+2{\bf C}_1}{4\Delta t})]^3\nonumber\\
&=\frac{[\frac{1}{N^2}+\frac{6}{N(N-1)}]\Sigma_0^3+[\frac{18}{N^2}+\frac{18N-30}{(N-1)^3}]\Sigma_0^2\Sigma_1+[\frac{6N-6}{N^3}+\frac{54N-90}{N(N-1)^2}]\Sigma_0\Sigma_1^2+[\frac{18N-30}{N^2(N-1)}+\frac{20N-44}{(N-1)^3}]\Sigma_1^3}{32\Delta t^3},
\end{align}
respectively.
\end{widetext}
We show in Appendix \ref{AppendixC}, that Eq.~\ref{DistD} reproduces Eq.~A11 for $k=2$ of Ref.~\cite{Vestergaard2014}.
Eq. \ref{EQ-35} is exact; it reproduces approximated solution Eq.~17 of Ref.~\cite{Vestergaard2014} to second order in $\frac{1}{N}$.

\begin{figure} 
	\includegraphics[scale=0.450]{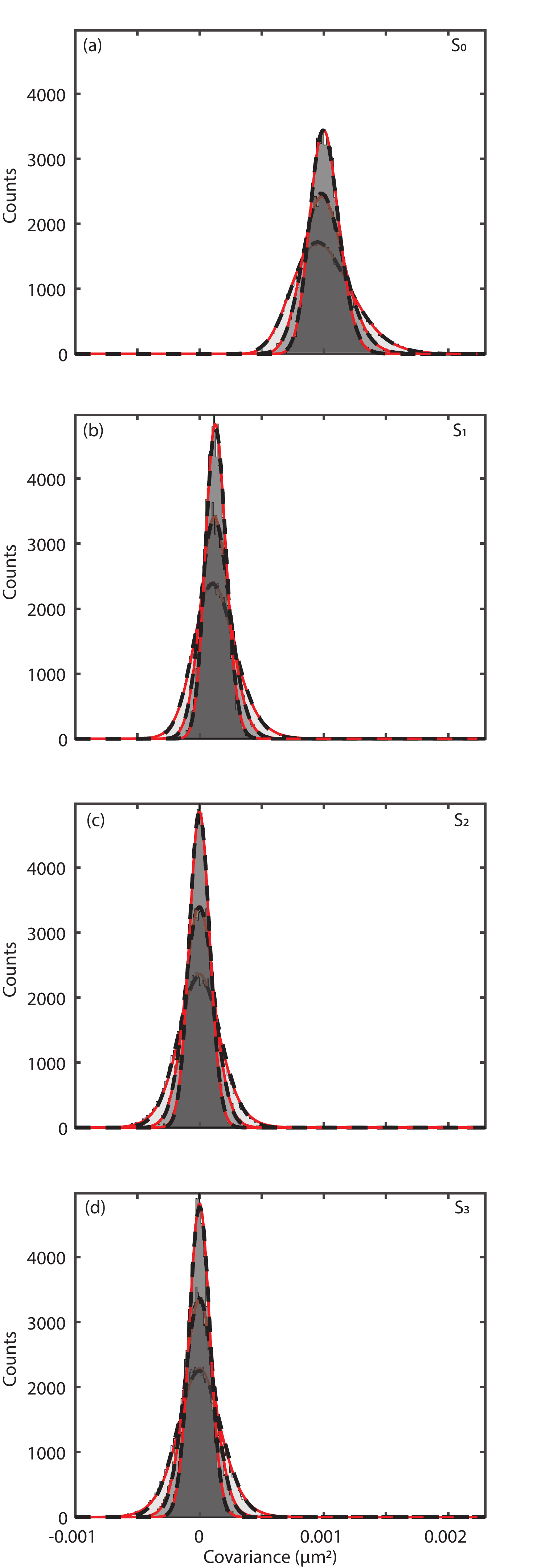}
	\caption{\label{fig:fig1_normal} Covariance distributions for simulated simple 2D diffusion for  $D = 0.0055~\mu$m$^2$s$^{-1}$, $\sigma^2 = 7.94\times 10^{-5}~\mu$m$^2$, and $\Delta t = \Delta t_E = 0.058s$. Distribution of covariances ${S}_0$ (a), ${S}_1$ (b), ${S}_2$ (c), and ${S}_3$ (d) for particle tracks of 19 (light gray), 39 (medium gray), and 79 (dark gray) steps are represented as histograms. Red lines correspond to the theoretical distributions. Black dashed lines correspond to the best fit of a skew normal distribution to the simulated distributions. With increasing number of time steps, the distribution narrows. Higher $S$ terms tend to center on 0.   }
\end{figure}

\begin{figure} 
\includegraphics[scale=0.450]{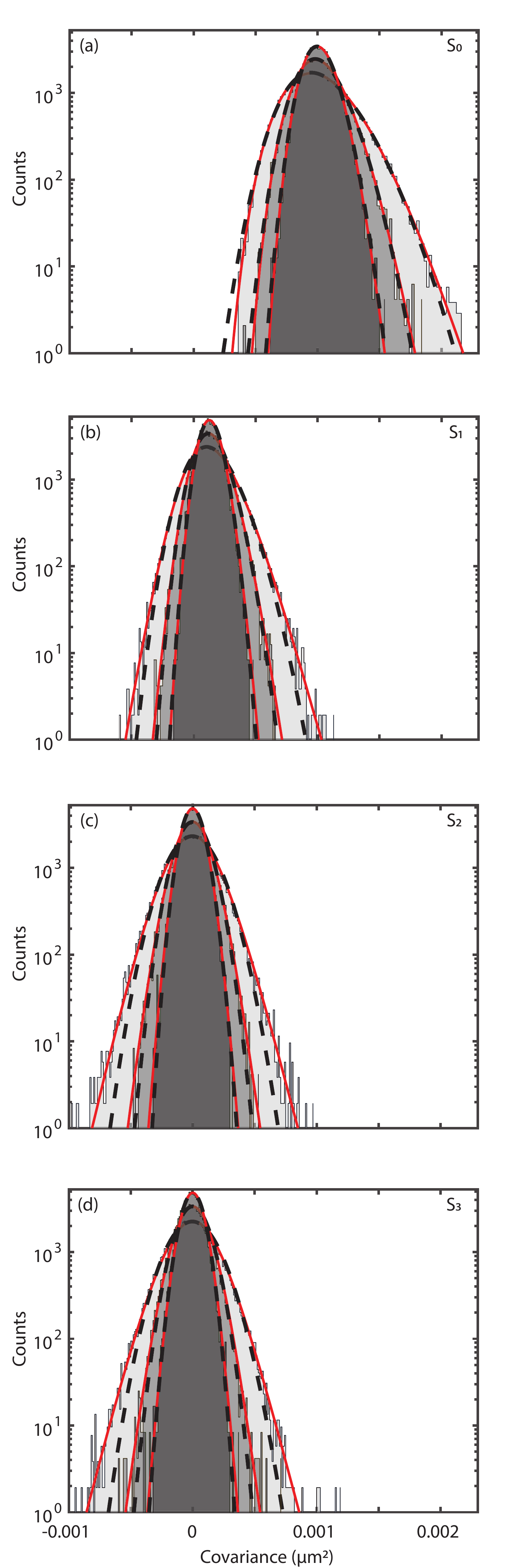}
\caption{\label{fig:fig1_normal_logy} Covariance distributions plotted on a logarithmic y-axis for simulated simple 2D diffusion, for the same data shown in FIG.~\ref{fig:fig1_normal}. Distribution of covariances ${S}_0$ (a), ${S}_1$ (b), ${S}_2$ (c), and ${S}_3$ (d) for particle tracks of 19 (light gray), 39 (medium gray), and 79 (dark gray) steps are represented as histograms. Red lines correspond to the theoretical distributions. Black dashed lines correspond to the best fit of a skew normal distribution to the simulated distributions. Only slight discrepancies between theory and skew normal curves are observed at the tails of the distributions.    }
\end{figure}

\section{Comparisons of theory to simulations and experiments}
\label{Sec:Comparisons}

\subsection{Simulations of simple 2D diffusion }
\label{sec:simpleD}

First, we compare the theory of Sec.~\ref{Sec:Theory} to simulated particle trajectories undergoing simple 2D diffusion,
generated as described in Ref.~\cite{Koo2016}. 
We simulated  $10^6$ steps, and set $D=0.0055~\mu$m$^2$ s$^{-1}$, 
$\sigma^2 = 7.94\times10^{-5}~\mu$m$^2$, and $\Delta t = \Delta t_E = 0.058s$. Calculating the $r^2$ value sets the scale for statistical quantities, where $r^2 = 4D\Delta t$ = $0.0013~\mu$m$^2$. The scale of the localization noise is then $\frac{\sigma^2}{r^2} = 0.062$. Lastly, the root-mean-square (RMS) step size for this simulation is $35.7$ nm. These values describing the datasets analyzed in this paper are shown together in Table~I. 
We partitioned  these displacements into tracks of varied length, $N$, ranging from 19 to 419 steps, and then calculated the 
 covariance matrix elements and diffusion coefficient (Eq.~\ref{EQ-28}) for each such track.

Normalized histograms of thus calculated $S_0$, $S_1$, $S_2$, and $S_3$  are plotted in Fig.~\ref{fig:fig1_normal}(a), (b), (c), and (d), respectively, for different track lengths $N=19, 39$, and $79$.
The $S_0$-distributions are entirely positive, as required.
The mean of the $S_1$-distributions are positive, as expected when motion blur dominates static localization error.
For $n>1$, the means of the $S_n$-distributions appear to be zero, also as expected. 
The distributions become progressively narrower with increasing track length, because a longer track length represents a more accurate measurement.
Also shown in Fig.~\ref{fig:fig1_normal} as the solid red lines are the corresponding theoretical distributions (Eq.~\ref{EQ-16}),
calculated using only the experimental mean covariance matrix elements.
Clearly, these theoretical predictions closely match the simulated distributions.
Logarithmic-linear plots of the same data and model are
presented in Fig.~\ref{fig:fig1_normal_logy}, demonstrating that the simulated and theoretical distributions continue to agree well, even in the far tails.

The Central Limit Theorem informs us that the distributions must each approach a Gaussian in the limit of large $N$.
However, the simulated and theoretical distributions for $N=19$, and even for $N=39$, are noticeably skewed.
To provide a simple way to empirically gauge the mean, variance, and third central moment of measured covariance distributions by fitting,
we approximate them with a skew normal distribution  \cite{Henze1986}:
\begin{equation} \label{eq:skew}
p(S_n) = 
\frac{1}{\sqrt{2\pi}\rho } 
e^ { - \frac{(S_n - \zeta)^2}{2 \rho^2 }  }
\left[  1 + {\rm erf} \left(   \frac{\alpha (S_n - \zeta)}{ \sqrt{2}\rho} \right) \right],
\end{equation}
with mean
\begin{equation}
\left < S_n \right > =\zeta +\sqrt{\frac{2}{\pi}} \frac{\alpha \rho}{\sqrt{1+\alpha}}
\end{equation}  
variance 
\begin{equation}
\sigma^2_{S_n} = \rho^2 (1 - \frac{2\alpha^2}{\pi}(1+\alpha) ) 
\end{equation}
and third central moment
\begin{equation}
 \left < (S_n-\left < S_n \right >)^3 \right > = \frac{4-\pi}{2}\frac{(\delta\sqrt{2/\pi})^3}{(1-2\delta^2/\pi)^\frac{3}{2}}(\rho^2 (1 - \frac{2\alpha^2}{\pi}(1+\alpha) ))^\frac{3}{2}
 \end{equation}
 
\begin{figure} 
	\includegraphics[scale=0.450]{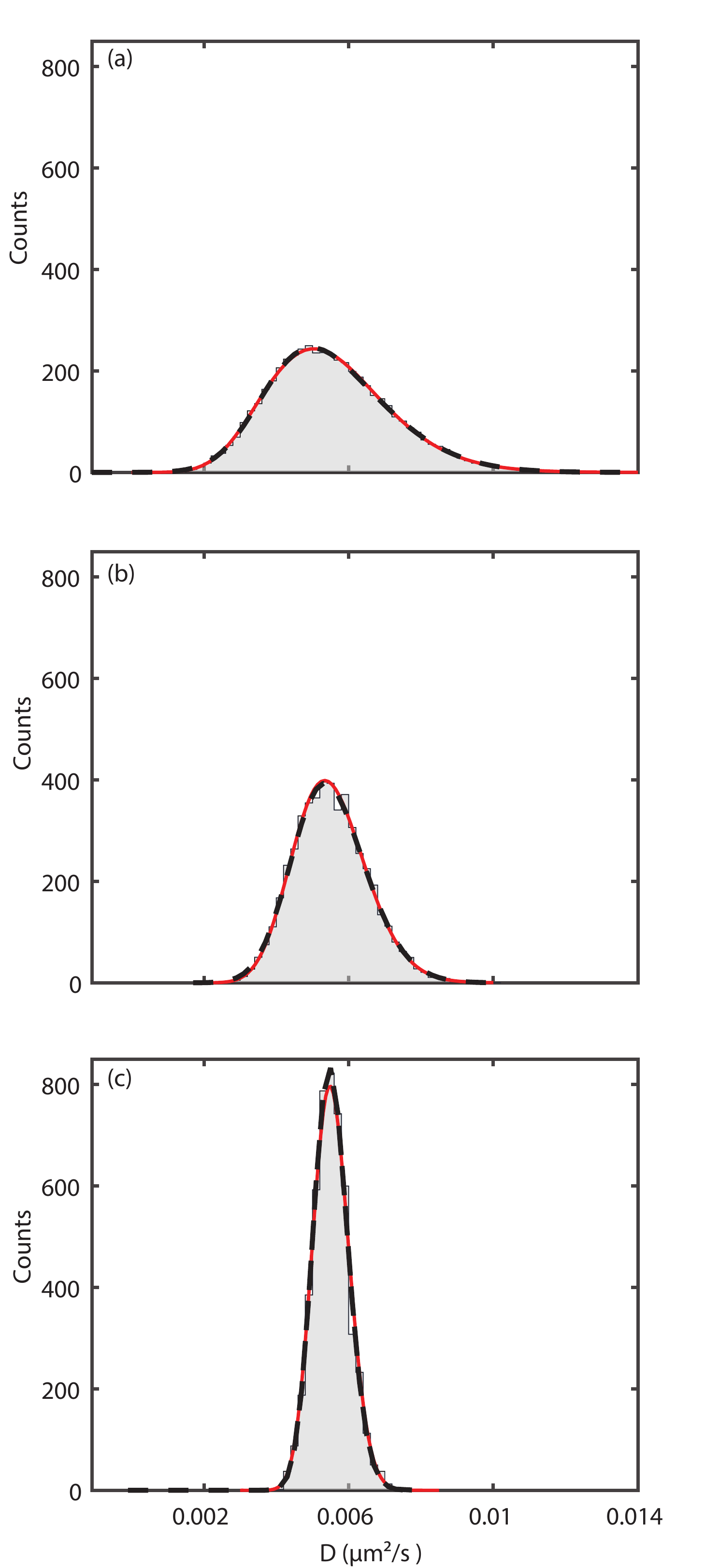}
	\caption{\label{fig:Dfig} Distribution of diffusion coefficients, $D$, for simulated simple 2D diffusion  with $D = 0.0055~\mu$m$^2$s$^{-1}$, $\sigma^2 = 7.94\times 10^{-5}~\mu$m$^2$, and $\Delta t = \Delta t_E = 0.058s$ (same data as shown in FIGs.~\ref{fig:fig1_normal}-\ref{fig:fig1_normal_logy}), for track lengths of (a) 29 steps, (b) 79 steps, and (c) 319 steps, shown using a linear $y$-axis. The red curve represents the theoretical prediction given by Eq.~\ref{DistD}. The black dashed curve corresponds to the best skew normal fit to the simulated distribution. 
	}  
\end{figure}

\begin{figure} 
	\includegraphics[scale=0.450]{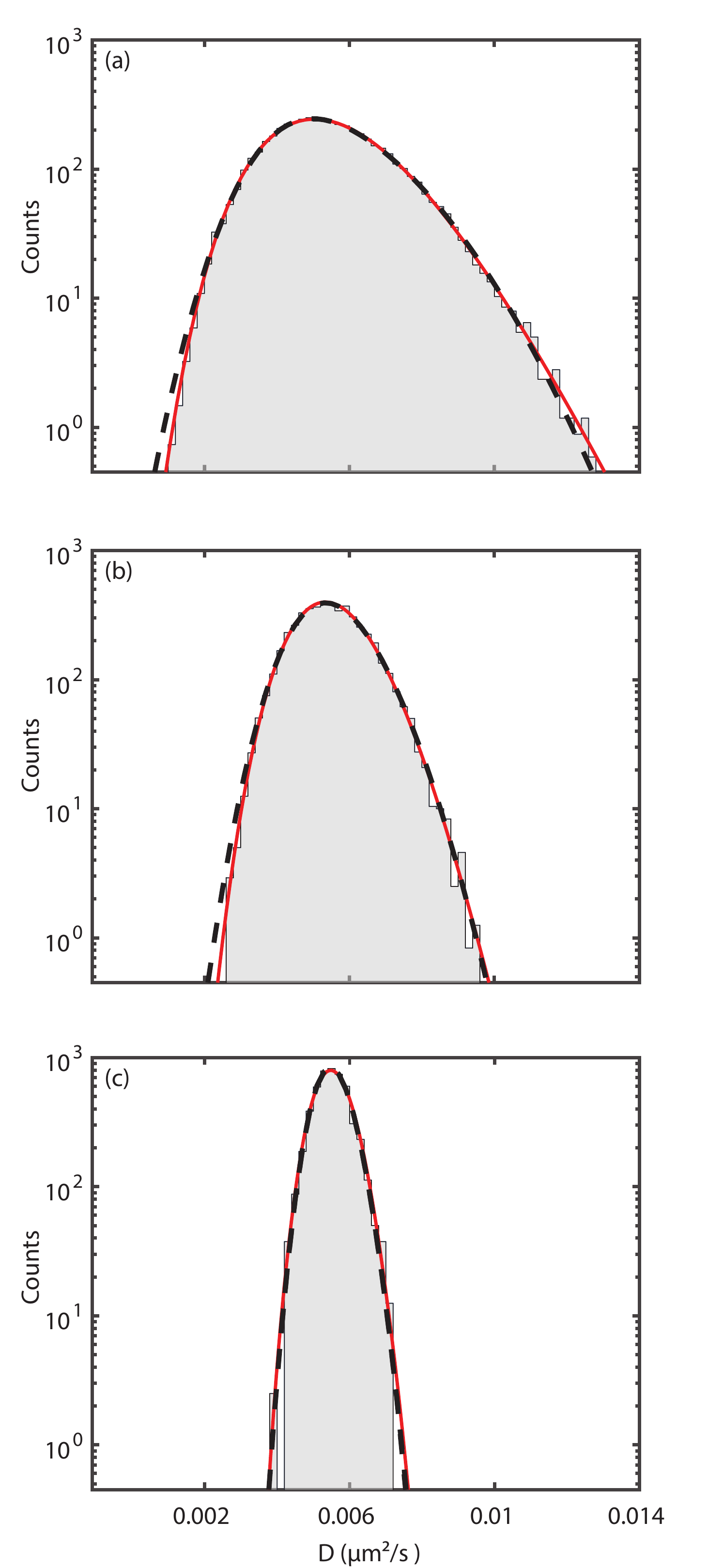}
	\caption{\label{fig:Dlogyfig} Distribution of diffusion coefficients, $D$, plotted on a logarithmic y axis (same data as FIG.~\ref{fig:Dfig}), for track lengths of (a) 29 steps, (b) 79 steps, and (c) 319 steps, shown using a logarithmic $y$-axis.
		The red curve represents the theoretical prediction given by Eq.~\ref{DistD}.
		The black dashed curve corresponds to the best skew normal fit to the simulated distribution. }
\end{figure}

The black lines in Figs.~\ref{fig:fig1_normal} and~\ref{fig:fig1_normal_logy} are the best fits of a skew normal distribution to distributions sampled from the simulations,
varying $\alpha$, $\zeta$, and $\rho$ as fitting parameters.
Near the peak, it is evident that the skew normal fit is able to accurately capture the shape of each distribution.
While the best-fit skew normal distribution shows small deviations from both theory and simulated data (Fig.~\ref{fig:fig1_normal_logy}) in the far tails, we judge that it provides a good approximation to both.
The distributions of estimated  diffusion coefficients are shown in Figs.~\ref{fig:Dfig} and~\ref{fig:Dlogyfig}, together with skew-normal fits. Again close examination reinforces that the theoretical curves very closely match the simulated distributions, and that the skew-normal fits provide an excellent description with only small deviations in the tails.



\begin{table*}
	\begin{tabular}{| c | c | c | c | c | c | c |}
		\hline
		\bf{Dataset} & \bf{D} & $\bf{\boldsymbol{\sigma^2}}$ \bf{($\bf{\boldsymbol{\mu} m^2}$)}  & $\bf{\Delta t}$ \bf{(s)} & $\bf{r^2}$ \bf{($\boldsymbol{\mu} \mathbf{m^2}$)} & \bf{$\boldsymbol{\sigma^2/r^2}$} & \bf{RMS step size ($\bf{nm}$)} \\ \hline
		Simulated simple 2D diffusion & 0.0055 ($\mu m^2/s$) & 7.94 $\times 10^{-5}$ & 0.058 & 0.00128 & 0.062 & 35.7 \\ \hline
		Simulated simple 2D diffusion & 0.1389 ($\mu m^2/s$) & 0.0015 & 0.01 & 0.0056 & 0.270 & 74.6 \\ \hline
		Simulated simple 2D diffusion & 0.0062 ($\mu m^2/s$) & 0.0013 & 0.01 & 2.48 $\times 10^{-4}$ & 5.24 & 15.7 \\ \hline
		Optically-trapped bead in water & 0.14 ($\mu m^2/s$) & 2.80 $\times 10^{-7}$ & 4.8 $\times 10^{-5}$ & 1.34 $\times 10^{-5}$ & 0.021 & 3.7 \\ \hline
		Optically-trapped bead in PEO & 3.3 $\times 10^{-5}$ ($\mu m^2/s^{0.20}$) & 2.80 $\times 10^{-7}$ & 5.0 $\times 10^{-5}$ &9.1 $\times 10^{-5}$ & 0.031 & 3.0 \\ \hline
		Gene locus \textit{mmf1} & 0.0061 ($\mu m^2/s^{0.39}$) & 0.0027 & 0.058 & 0.0080 & 0.338 & 89.4 \\ \hline
		Simulated fBm & 0.0055 ($\mu m^2/s^{0.44}$) & 0.0032 & 0.058 & 0.0063 & 0.51 & 79.4 \\ \hline
	\end{tabular}
	\caption{\label{table_params} Table of the SPT parameters and calculated length scale values for the datasets analyzed in this paper. For 2D simple diffusion the length scale of the step size is $r^2 = 4D\Delta t$. For optical tweezers data, which is one-dimensional, $r^2 = 2D\Delta t$ for the bead in water. For the bead in PEO solution, which exhibits fractional Brownian motion (fBm)  $r^2 = 2D\Delta t^{\alpha}$, where $\alpha$ is the anomalous exponent. For the gene locus and simulated fBm tracks, $r^2 = 4D\Delta t^{\alpha}$. }
\end{table*}

\begin{figure}
	\includegraphics[scale=0.43]{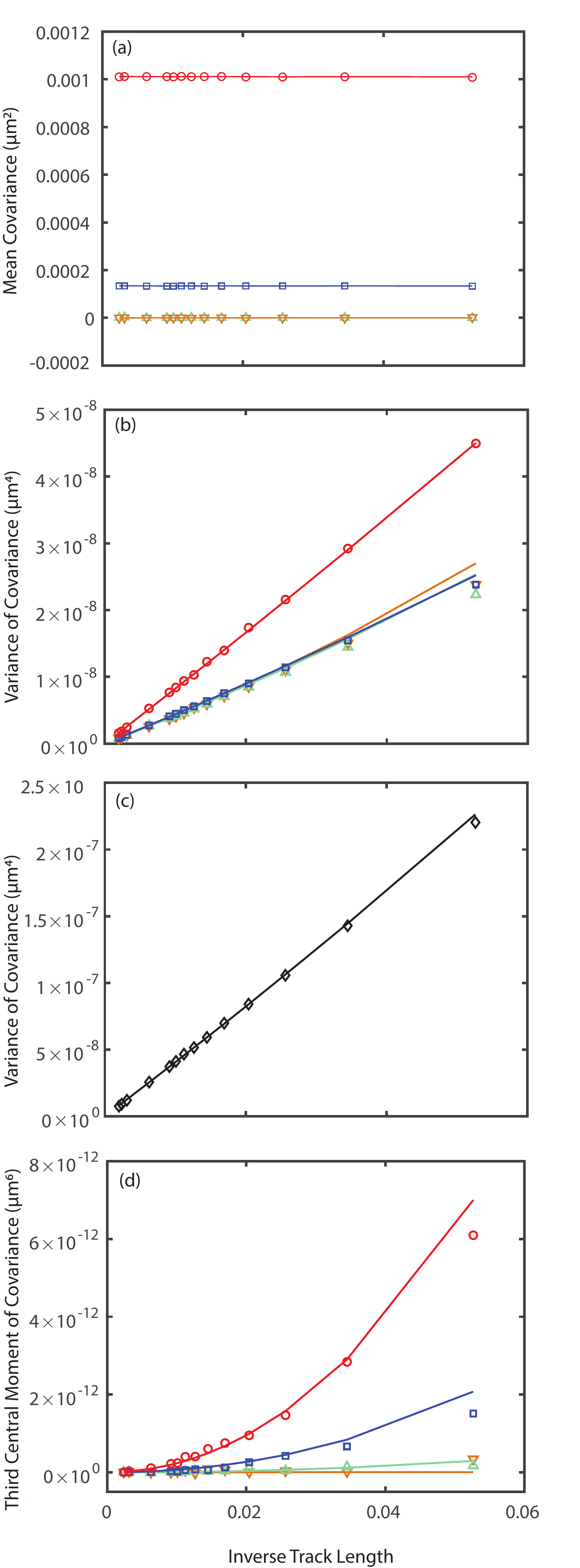}
	\caption{\label{fig:fig2_normal} Dependence of the covariance matrix estimates on the track length for simple 2D diffusion data (same data as FIGs.~\ref{fig:fig1_normal}-\ref{fig:Dlogyfig}). (a) The means of the covariances $S_0$ (red circle),  $S_1$ (blue square),  $S_2$ (green triangle), and  $S_3$ (orange upside-down triangle) are independent of track length. The maximum y-value corresponds to $0.94$ in units of the mean-square step size ($0.00128$ $\mu$m$^2$).
(b) Variance of the covariance {\em vs.} inverse track length for ${S}_0$ (red circle), ${S}_1$ (blue square),  ${S}_2$ (green triangle), and ${S}_3$ (orange upside-down triangle). 
Theory (Eq.~\ref{Variance2D}) is shown as the solid lines. Maximum y-value corresponds to $0.031$ in units of the mean-square step size ($0.00128$ $\mu$m$^4$)
(c) Variance of the covariance of ${S}_0 + 2{S}_1$ {\em vs.} inverse track length.
Theory ($\frac{{\rm Tr}({\bf \Sigma} ({\bf C}_0+2{\bf C}_1))^2}{4}$) is shown as the solid line. Maximum y-value corresponds to $0.15$ in units of the mean-square step size ($0.00128$ $\mu$m$^4$)
(d) Third central moment of the covariance {\em vs.} inverse track length.
In this case too, theory (Eq.~\ref{Skewness2D}) agrees well with the results of simulations. Maximum y-value corresponds to $0.0038$ in units of the mean-square step size ($0.00128$ $\mu$m$^6$)
} 
\end{figure}

Fig.~\ref{fig:fig2_normal} shows how estimates of the covariance distributions from skew-normal fits depend on the inverse length of the tracks. Estimated mean (Fig.~\ref{fig:fig2_normal}a), variance (Fig.~\ref{fig:fig2_normal}b) and 3rd moment (Fig.~\ref{fig:fig2_normal}d) for ${S}_0$, ${S}_1$, ${S}_2$, and ${S}_3$ are compared to the ground-truth mean and to theoretical predictions for the variance and 3rd moment (Eqs. \ref{mean}-\ref{Skewness2D}). Because the covariance distributions are skewed, we include the third central moment to compare the fitted third moment (related to the skewness, see Eq.~\ref{skewnessa12}) to the theoretically predicted. In Fig.~\ref{fig:fig2_normal}(c), we show specifically the variance of $S_0 + 2S_1$. This combination removes static localization noise, leaving a result proportional to the diffusion
coefficient.
In every case, the best fit parameters match closely the corresponding theoretical values,
supporting the utility of the skew normal function as a simple route to describe covariance and diffusivity distributions. 

\begin{figure}
	\includegraphics[scale=0.5]{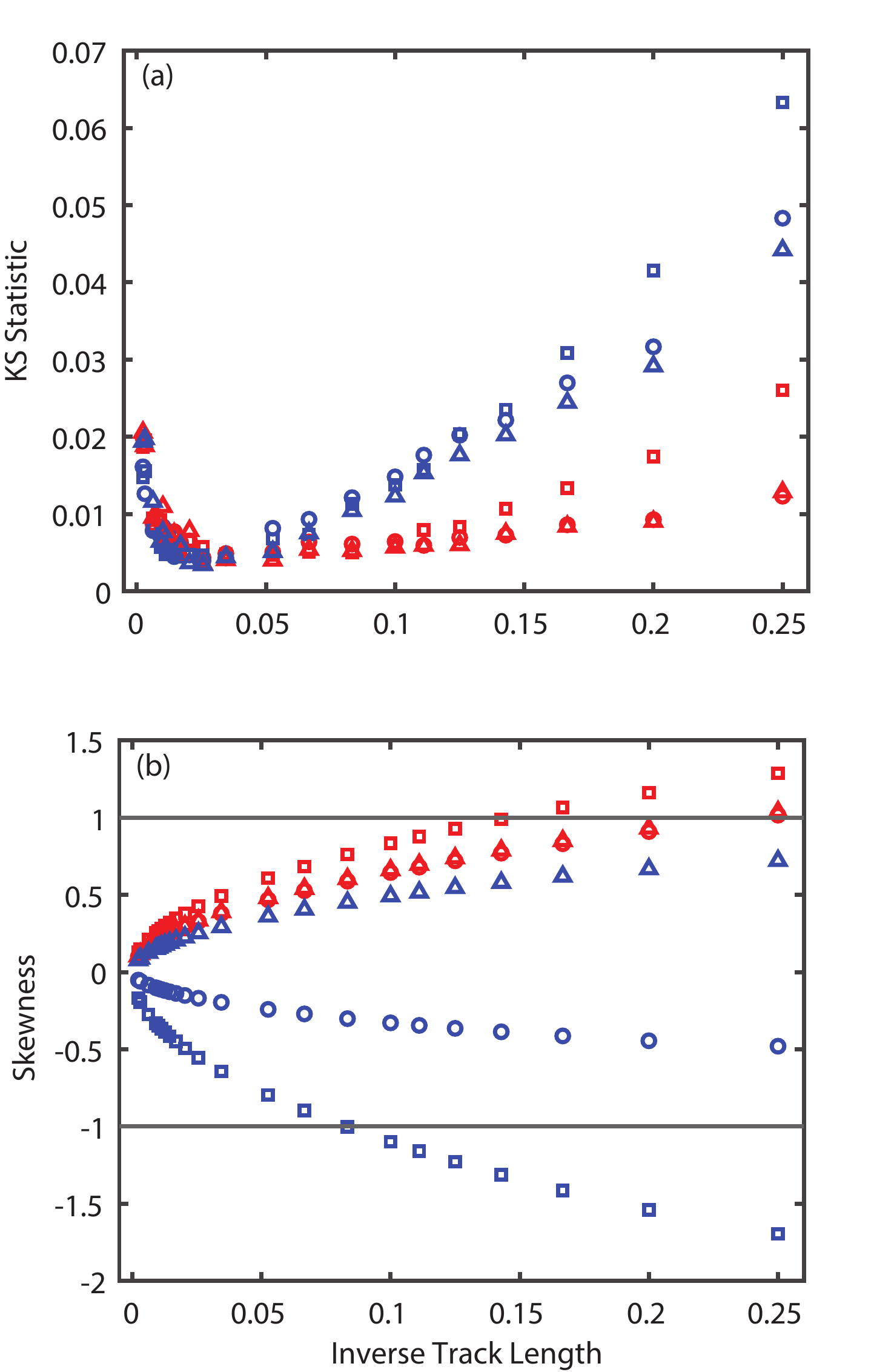}
	\caption{\label{fig:ks_skew} (a) KS statistic on fits of $S_0$ (red) and $S_1$ (blue) versus inverse track length, for  $D = 0.0055~\mu$m$^2$s$^{-1}$, $\sigma^2 = 7.94\times 10^{-5}~\mu$m$^2$, and $\Delta t = \Delta t_E = 0.058s$  (triangles); $D = 0.1389~\mu$m$^2$s$^{-1}$, $\sigma^2 = 0.0015~\mu$m$^2$, and $\Delta t = \Delta t_E = 0.01s$  (circles); and $D = 0.0062~\mu$m$^2$s$^{-1}$, $\sigma^2 = 0.0013~\mu$m$^2$, and $\Delta t = \Delta t_E = 0.01s$ (squares). For very small track length, the KS statistic is large. (b) Corresponding theoretical skewness of $S_0$ and $S_1$ versus inverse track length. Gray solid lines signify skewness values of $\pm{1}$, the max/min skewness values that describe the skew normal distribution. For smaller track lengths, the skewness approaches $\pm{1}$. 
	} 
\end{figure}

 We further evaluate whether a skew normal distribution provides a good approximation to the data by applying the Kolmogorov-Smirnov (KS) test, which uses the difference between the CDFs of two distributions to estimate whether they originate from the same underlying distribution \cite{Massey1951}. We performed analysis with  track lengths $N = 4-419$. 
 A plot of the KS statistic versus inverse track length for fits of $S_0$ and $S_1$ covariance distributions on simple 2D diffusion simulations, with  $D = 0.1389~\mu$m$^2$s$^{-1}$, $\sigma^2 = 0.0015~\mu$m$^2$, and $\Delta t = \Delta t_E = 0.01s$ ($r^2$ = $0.0056~\mu$m$^2$, $\frac{\sigma^2}{r^2} = 0.27$, RMS step size = $74.6$ nm), and $D = 0.0062~\mu$m$^2$s$^{-1}$, $\sigma^2 = 0.0013~\mu$m$^2$, and $\Delta t = \Delta t_E = 0.01s$ ($r^2$ = $0.00025~\mu$m$^2$, $\frac{\sigma^2}{r^2} = 5.24$, RMS step size = $15.7$ nm), as well as the diffusivity parameters used in Figs.~\ref{fig:Dfig}-\ref{fig:fig2_normal}, is shown in Fig.~\ref{fig:ks_skew}(a). For smaller track lengths (i.e. $N$=19-49), the KS statistic is small, indicating a better fit, relative to other track lengths. Interestingly, very short track lengths (less than about 19) produce a poor fit using the skew normal distribution. This phenomenon can be explained by a limitation of the skew normal distribution, which is constrained to a maximum/minimum skewness of $\pm{1}$. For the same three data sets, Fig.~\ref{fig:ks_skew}(b) shows the theoretically calculated skewness values for $S_0$ and $S_1$ versus inverse track length. For very small track lengths, the skewness can go beyond
the range between
$\pm{1}$. In the case of $D = 0.0062~\mu$m$^2$s$^{-1}$, for example, the skewness of $S_1$ is
more negative than $-1$ at N = 12, rendering the skew normal distribution a poor descriptor of the data for this and smaller track lengths. To avoid this limitation, we use the exact theory to calculate the covariance distribution for very small track lengths, instead of fitting with the skew normal distribution. 

\subsection{\label{sec:level5} Confined diffusion of optically-trapped beads}
\label{sec:Beads}
\subsubsection{Optically-trapped bead in water}
\label{BeadInWater}
Next, we compare the theory of Sec.~\ref{Sec:Theory} to measurements of confined diffusion of the $x$-coordinate of a $1~\mu$m-diameter optically-trapped polystyrene bead suspended in water, from Ref.~\cite{Yan2017}.
Specifically,  ten independent time series of the bead position were recorded at 20~kHz,  each containing 100,000 points, using a National Instruments DAQ PCIe-6343.
In an optical trap, a particle experiences thermal Brownian motion, subject to a quadratic confining potential \cite{Ashkin1987}.
Theoretical results for the mean-square displacement (MSD) and covariance matrix elements of such a particle are presented in Appendix \ref{AppendixD},
demonstrating that the covariance matrix elements are non-zero for $n>1$, in contrast to the case of simple diffusion.

\begin{figure} 
	\includegraphics[scale=0.47]{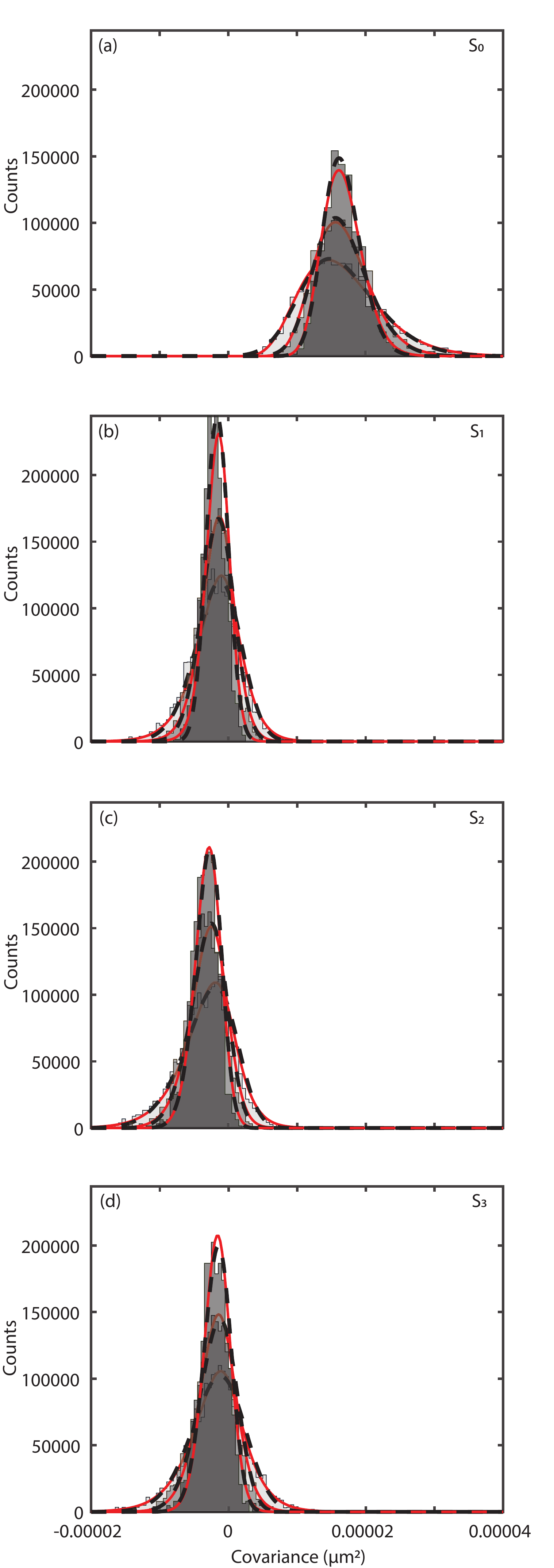}
	\caption{\label{fig:fig1_OTbead} Covariance distributions for a bead in water suspended in an optical trap, measured by optical tweezers.  Covariances ${S}_0$ (a), ${S}_1$ (b), ${S}_2$ (c), and ${S}_3$ (d) for particle tracks of 19 (light gray), 39 (medium gray), and 79 (dark gray) steps are represented as histograms. Red lines correspond to the theoretical distributions. Black dashed lines correspond to the best fit of a skew normal distribution to the simulated distributions.
	 }
\end{figure}

\begin{figure} 
	\includegraphics[scale=0.46]{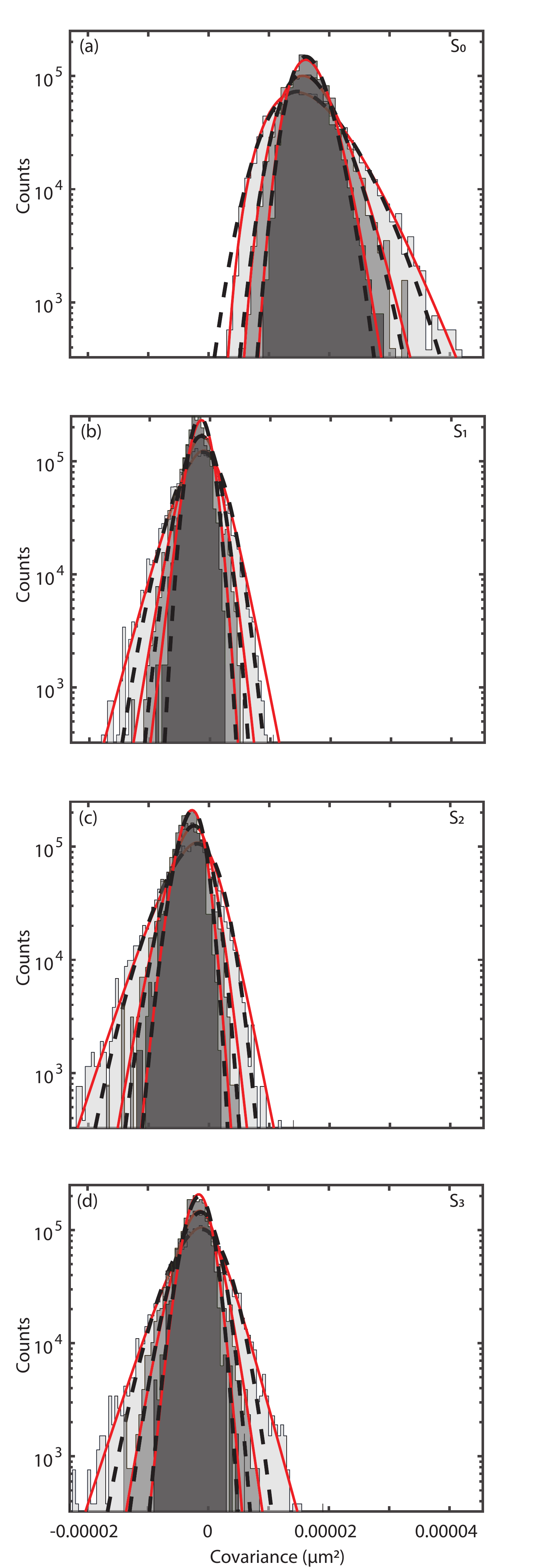}
	\caption{\label{fig:fig1_OTbead_logy} Covariance distributions for a bead in water suspended in an optical trap, measured by optical tweezers, plotted on a logarithmic y-axis. Distribution of covariances ${S}_0$ (a), ${S}_1$ (b), ${S}_2$ (c), and ${S}_3$ (d) for particle tracks of 19 (light gray), 39 (medium gray), and 79 (dark gray) steps are represented as histograms. Red lines correspond to the theoretical distributions. Black dashed lines correspond to the best fit of a skew normal distribution to the simulated distributions. 
}
\end{figure}

\begin{figure}
	\includegraphics[scale=0.49]{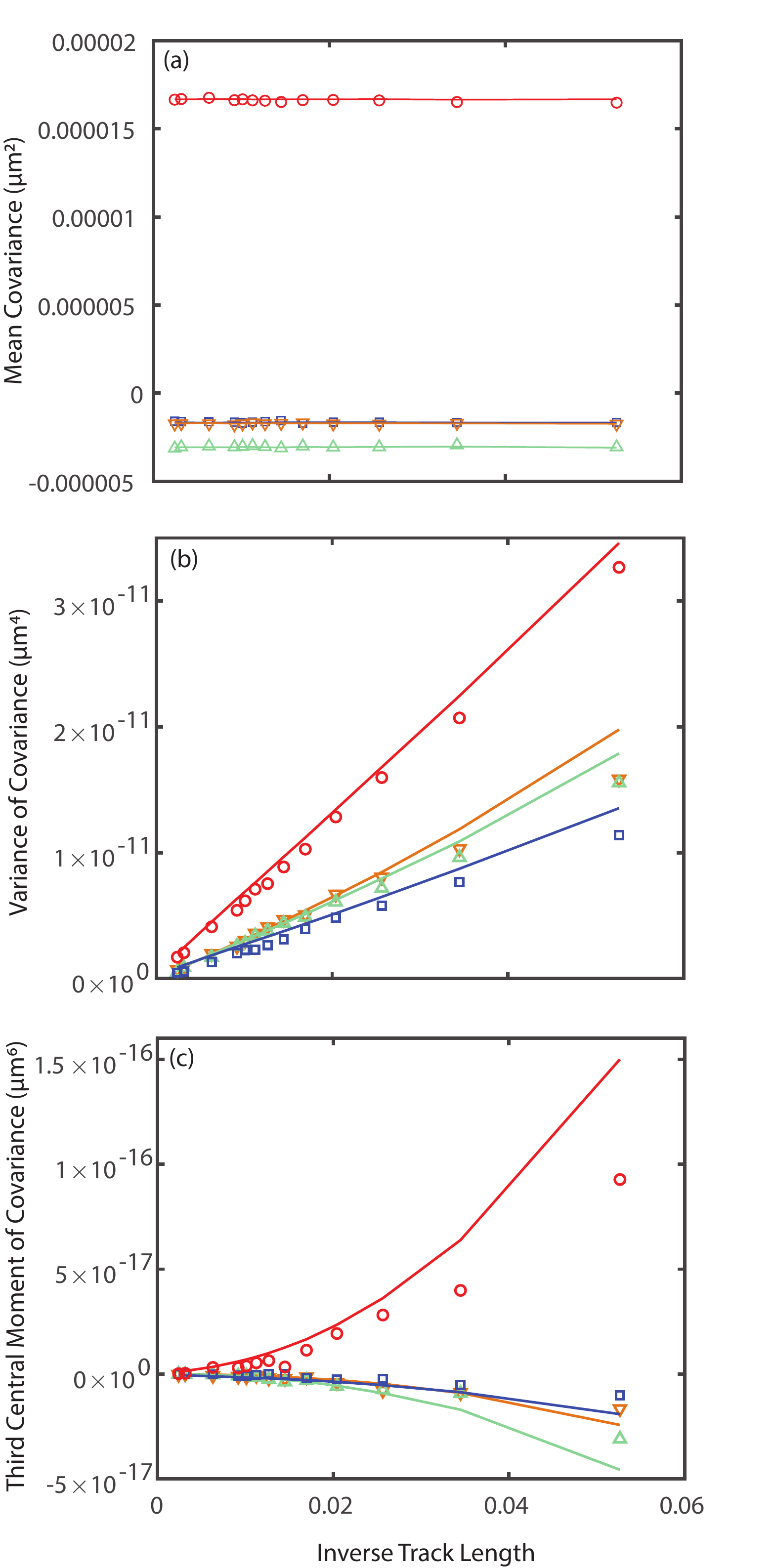}
	\caption{\label{fig:fig2_free}  Dependence of the covariance matrix estimates on the track length for a bead in water suspended in an optical trap, measured by optical tweezers. (a) The mean of the covariances $S_0$ (red circle),  $S_1$ (blue square),  $S_2$ (green triangle), and  $S_3$ (orange upside-down triangle). Theory (straight lines) is calculated by Eq.~\ref{EQ7}. Maximum y-value corresponds to $1.49$ in units of the mean-square step size ($1.34\times 10^{-5}$ $\mu$m$^2$). (b) Variance of the covariance vs. inverse track size of ${S}_0$ (red circle), ${S}_1$ (blue square),  ${S}_2$ (green triangle), and ${S}_3$ (orange upside-down triangle). The theoretical variances (solid lines) agree well with the data. The theory is calculated by Eq.~\ref{Variance1D}. Maximum y-value corresponds to $0.18$ in units of the mean-square step size ($1.34\times 10^{-5}$ $\mu$m$^4$)(c) Third central moment of the covariance vs. inverse track length. The theoretical values (solid lines) are calculated using Eq.~\ref{Skewness1D}. Maximum y-value corresponds to $0.062$ in units of the mean-square step size ($1.34\times 10^{-5}$ $\mu$m$^6$) } 
\end{figure}

%
%

Normalized histograms of the experimental covariance distributions for $S_0$, $S_1$, $S_2$, and $S_3$  are shown in Fig.~\ref{fig:fig1_OTbead},  along with the corresponding theoretical curves (red lines,  Eq.~\ref{Variance2D}) and best fits to the skew normal distribution (black lines). This figure demonstrates good agreement between theory and experiment, and that the skew normal fit describes both well.
The same plots are presented on logarithmic-linear axes in Fig.~\ref{fig:fig1_OTbead_logy}, again revealing that the skew normal fits show only small discrepancies in the tails of the distributions. Results from the skew normal  fits  are compared to theory in Fig.~\ref{fig:fig2_free}, again showing good agreement.

\begin{figure} 
	\includegraphics[scale=0.45]{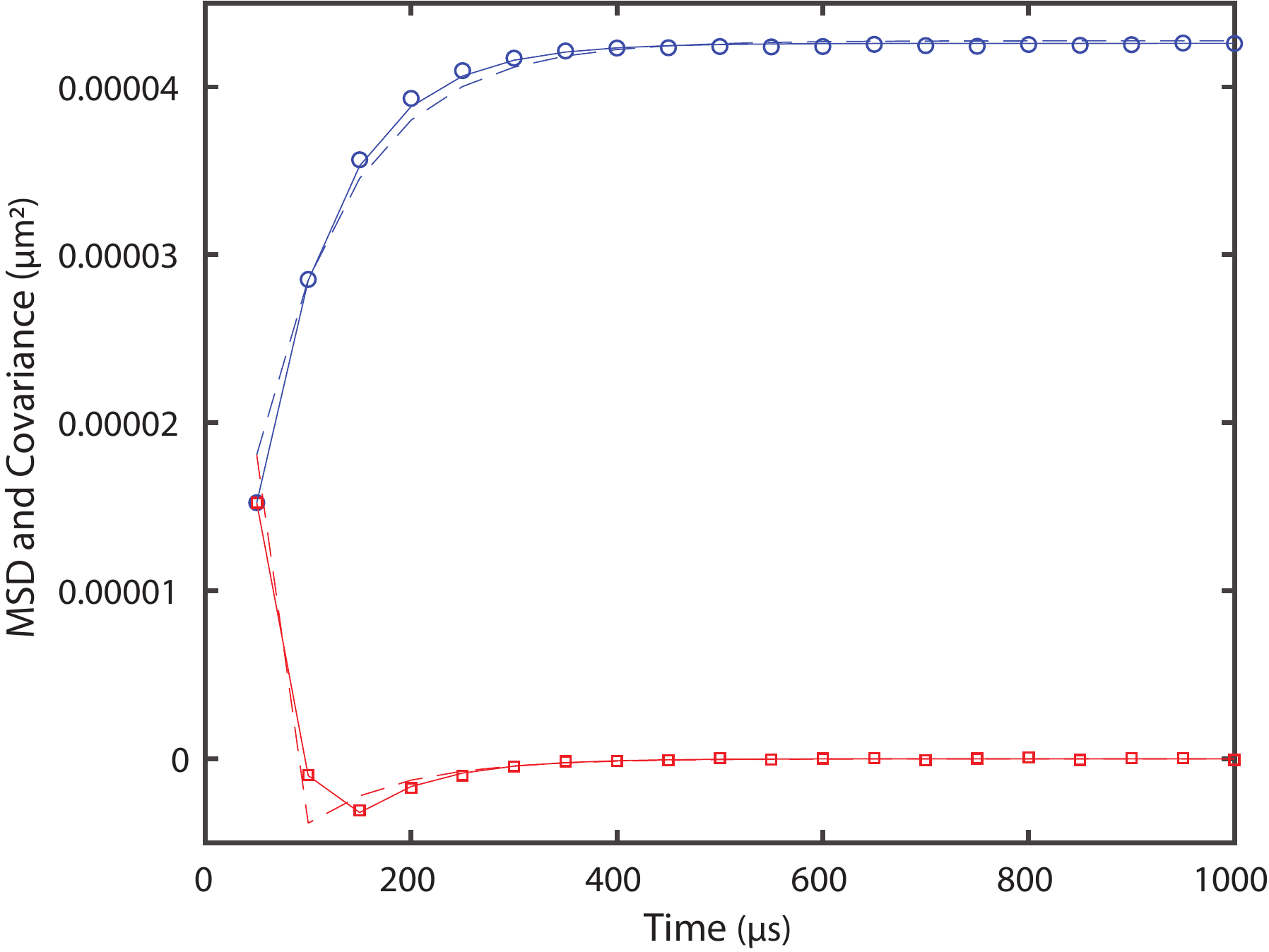}
	\caption{\label{fig:OTfit} Experimental MSD (circles) and covariance (squares) versus time delay for an optically-trapped bead in water,
	compared to corresponding best-fit model MSDs (blue solid and dashed lines) and covariances (red solid and dashed lines).
	The  experimental MSD and covariance were determined from 12650 79-step tracks.
	Errors (standard errors of the mean) are smaller than the plotted points, and therefore no error bars are plotted.
	The MSD and the covariance were fit using Eqs.~\ref{TweezersMSD} through \ref{TweezersSigman} by varying $\kappa$,  $\tau$, and $\sigma^2$ and by varying $\Delta t_E$ (solid line) or by fixing $\Delta t_E = 0$ (dashed line). Shown are the best fit curves with  values of $\kappa=0.160\pm0.008~{\rm pN~nm^{-1}}$, $\tau=0.078\pm0.005~{\rm ms}$, $\sigma^2=2.8\pm1.6\times 10^{-7}~{\rm~\mu m^2}$
	and $\Delta t_E=0.048\pm0.001~{\rm ms}$ ($\chi^2=23$) and  $\kappa=0.19~{\rm pN~nm^{-1}}$, $\tau=0.091~{\rm ms}$, and $\sigma^2=1.0\times 10^{-10}~{\rm~\mu m^2}$
	($\chi^2= 4.6\times 10^3$), for varied $\Delta t_E$ and $\Delta t_E = 0$, respectively. The former corresponds to  $r^2$ = $1.34\times 10^{-5}~\mu$m$^2$, $\frac{\sigma^2}{r^2} = 0.021$, and RMS step size = $3.7$ nm}
\end{figure}

The mean covariance (red) and mean MSD (blue) of the bead determined from 12650 79-step tracks for the first twenty time delays (i.e., $n=1...20$) is shown in Fig.~\ref{fig:OTfit}. The MSD is equivalent, via Fourier transformation, to the power spectral density (PSD), which is usually fit to determine  
the stiffness of the optical trap. However, the fitting is typically carried out without considering the integration time of the optical detector. To assess accuracy of such an approach, we fit the measured MSD and covariance simultaneously by varying $\kappa$,  $\tau$, and $\sigma^2$, along with $\Delta t_E$ (i.e., considering integration time) or by fixing  $\Delta t_E = 0$ (i.e., neglecting integration time). The results are shown as dashed ($\Delta t_E = 0$)  and solid (varied  $\Delta t_E$) lines in Fig.~\ref{fig:OTfit}. Including an integration time provides a significantly improved description of the data as judged by $\chi^2$ values (4600 vs 23 for $\Delta t_E = 0$ and varied $\Delta t_E$, respectively) and by visual inspection (Fig.~\ref{fig:OTfit}). Importantly, the best value of $\Delta t_E$ is 0.048 ms, just slightly smaller than the time between successive measurements (0.050~ms). This analysis suggests that it is important to
take into consideration a non-zero exposure time and the concomitant motion blur, when calibrating and analyzing optical tweezer data, as previously pointed out in Ref.~\cite{Wong2006}. Although Eq. \ref{TweezersMSD} reveals that the exponential time dependence of the MSD is unchanged by motion blur,
this circumstance is peculiar to this particular (exponential) form of the MSD \cite{Jakeman1974}.
In general, the shapes of the MSD and PSD are modified by motion blur (Appendix~\ref{AppendixE}).

\subsubsection{Optically-trapped bead in a viscoelastic PEO solution}
\label{BeadInPEO}
We also compared the theory of Sec.~\ref{Sec:Theory} to other data from Ref.~\cite{Yan2017}, 
comprising $10^6$ measurements of the $x$-coordinate of a $1~\mu$m-diameter optically-trapped polystyrene bead suspended
in  a 1.5~wt\% viscoelastic solution of long-chain (8~MDa) polyethylene oxide (PEO). 
Again,  ten independent time series were recorded at 20~kHz,  each containing 100,000 points, using a National Instruments DAQ PCIe-6343.
As shown in Ref.~\cite{Yan2017}, the PSD of a bead within this solution is well-described by a power law versus frequency,
suggesting that a power law in time could be an appropriate basis for
modeling the corresponding MSD and covariance,
 {\em i.e.} suggesting that the bead undergoes fractional Brownian motion (fBM), characterized by a subdiffusive exponent $\alpha <1$.
In Appendix~\ref{AppendixE}, we derive theoretical expressions for the MSD and covariance for a bead undergoing fractional Brownian motion (fBM) with
a subdiffusive exponent ($\alpha <1$), in the presence of motion blur and static localization noise.
In this case too, the covariance is non-zero for $n>1$.

\begin{figure} 
	\includegraphics[scale=0.450]{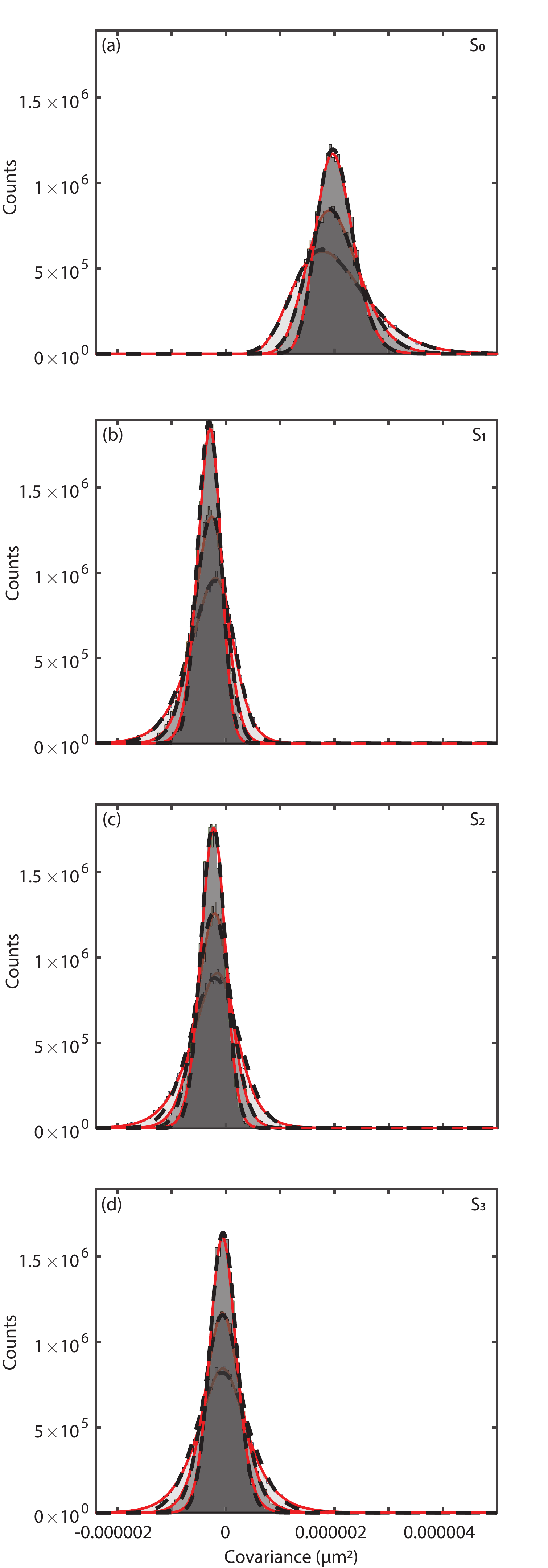}
	\caption{\label{fig:fig1_PEO} Covariance distributions for an optically-trapped bead in a 1.5~wt\% viscoelastic PEO solution. Covariances ${S}_0$ (a), ${S}_1$ (b), ${S}_2$ (c), and ${S}_3$ (d) for particle tracks of 19 (light gray), 39 (medium gray), and 79 (dark gray) steps are represented as histograms. Red lines correspond to the theoretical distributions. Black dashed lines correspond to the best fit of a skew normal distribution to the simulated distributions. With increasing number of time steps, the distribution narrows.  }
\end{figure}

\begin{figure} 
	\includegraphics[scale=0.450]{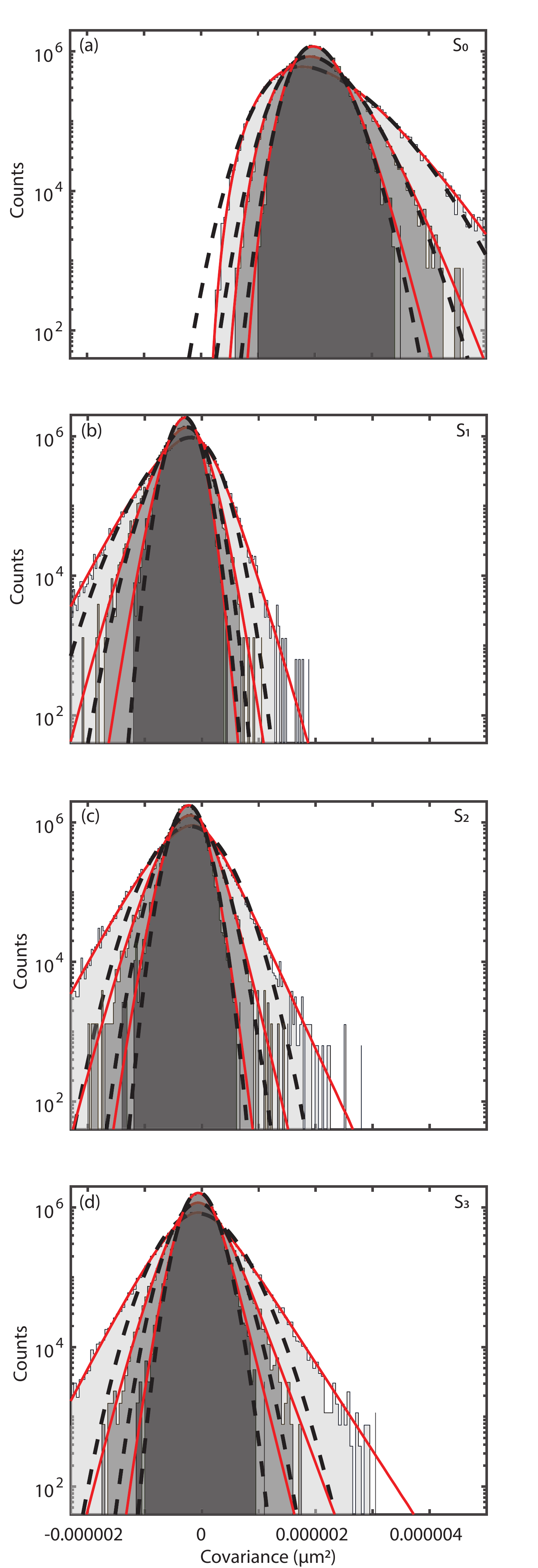}
	\caption{\label{fig:fig1_PEO_logy} Covariance distributions for an optically-trapped bead in a 1.5~wt\% viscoelastic PEO solution, plotted on a logarithmic y-axis. Distribution of covariances ${S}_0$ (a), ${S}_1$ (b), ${S}_2$ (c), and ${S}_3$ (d) for particle tracks of 19 (light gray), 39 (medium gray), and 79 (dark gray) steps are represented as histograms. Red lines correspond to the theoretical distributions. Black dashed lines correspond to the best fit of a skew normal distribution to the simulated distributions. Only slight discrepancies between theory and skew normal curves are observed at the tails of the distributions.    }
\end{figure}

\begin{figure}
	\includegraphics[scale=0.49]{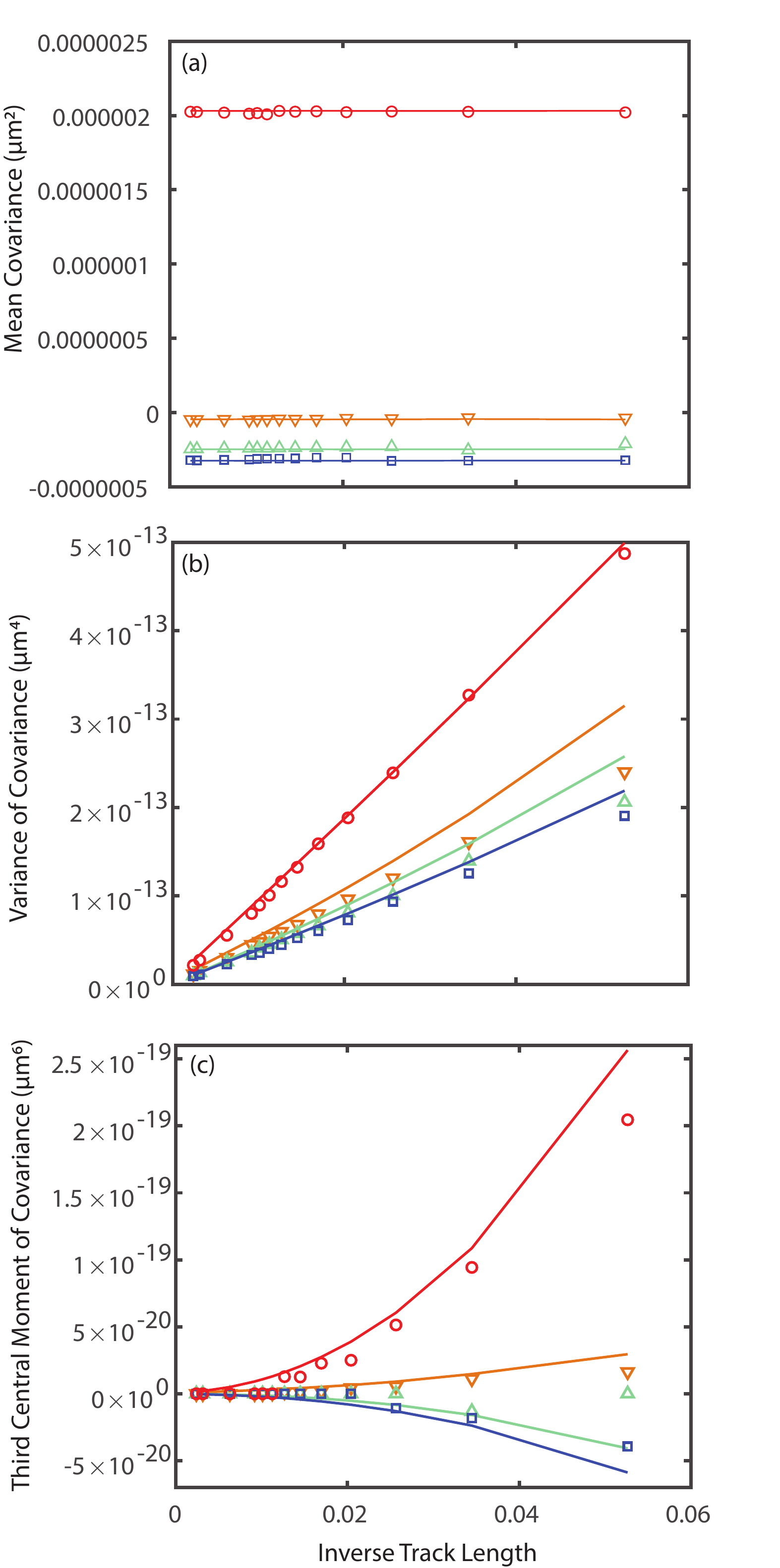}
	\caption{\label{fig:fig2_PEO} Dependence of the covariance matrix estimates on the track length for a bead in viscous PEO solution measured by optical tweezers. (a) The mean of the covariances $S_0$ (red circle),  $S_1$ (blue square),  $S_2$ (green triangle), and  $S_3$ (orange upside-down triangle). Theory (straight lines) is calculated by Eq.~\ref{EQ7}. Maximum y-value corresponds to $0.27$ in units of the mean-square step size ($9.1\times 10^{-6}$ $\mu$m$^2$) (b) Variance of the covariance vs. inverse track size of ${S}_0$ (red circle), ${S}_1$ (blue square),  ${S}_2$ (green triangle), and ${S}_3$ (orange upside-down triangle). The theoretical variances (solid lines) agree well with the data. The theory is calculated by Eq.~\ref{Variance1D}. Maximum y-value corresponds to $0.006$ in units of the mean-square step size ($9.1\times 10^{-6}$ $\mu$m$^4$) (c) Third central moment of the covariances $S_0$ (red circle),  $S_1$ (blue square),  $S_2$ (green triangle), and  $S_3$ (orange upside-down triangle) vs. inverse track length. The theoretical values (solid lines) are calculated using Eq.~\ref{Skewness1D}. Maximum y-value corresponds to $3.32\times 10^{-4}$ in units of the mean-square step size ($9.1\times 10^{-6}$ $\mu$m$^6$).  } 
\end{figure}

Figs.~\ref{fig:fig1_PEO} and \ref{fig:fig1_PEO_logy} show
the experimental covariance distributions, represented as histograms,
overlaid with the theoretical distributions from Eq.~\ref{1DProbabilityDensity} (red lines) and best-fit skew normal  curves (black lines).
Fig.~\ref{fig:fig2_PEO} displays the skew normal fitting results versus track length, together with the correspponding theoretical predictions.
For all of these figures, there is good agreement between theory and experiment.

\begin{figure} 
	\includegraphics[scale=0.45]{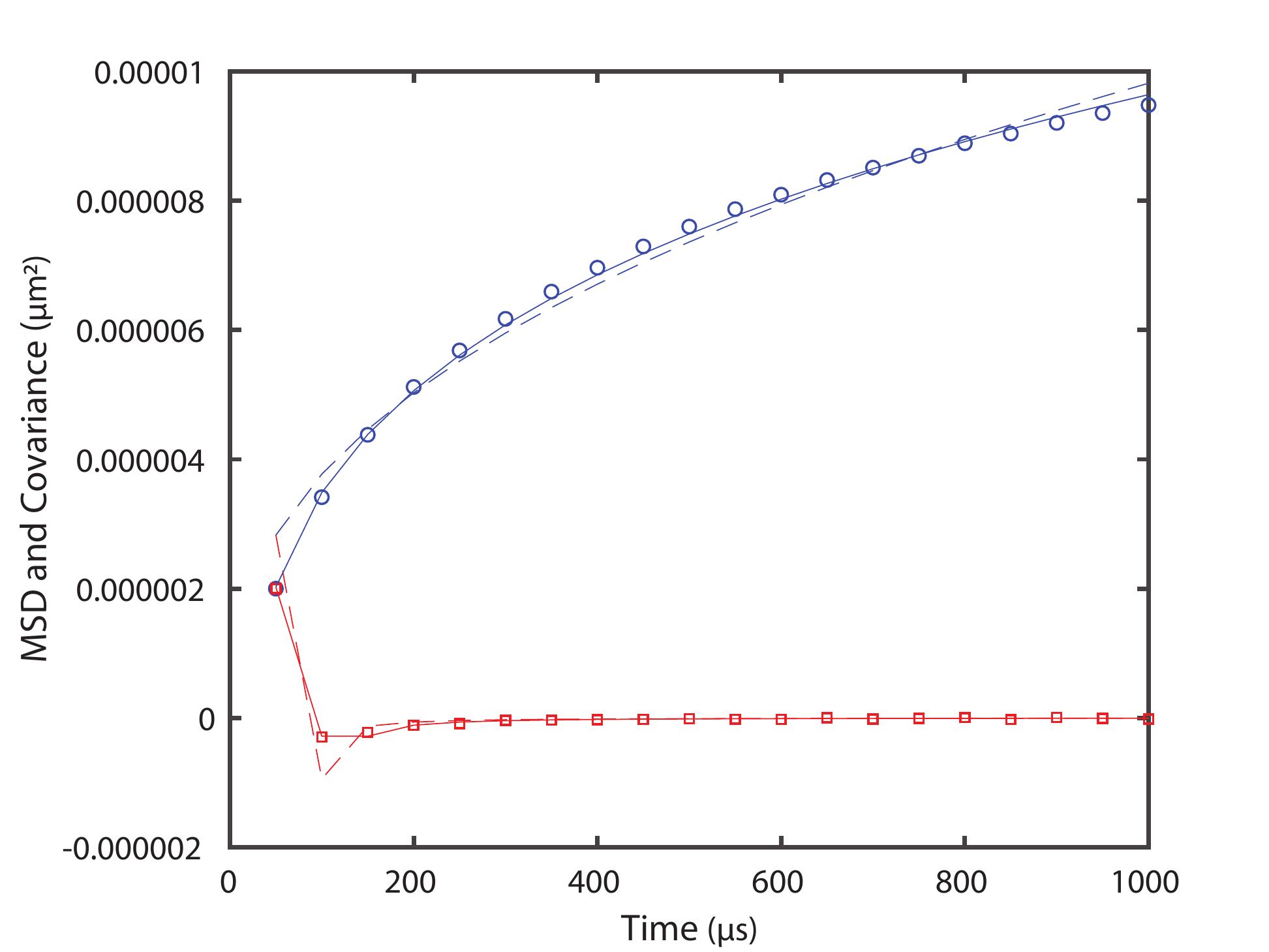}
	\caption{\label{fig:OTfit_PEO} Experimental MSD (circles) and covariance (squares) versus time delay for an optically-trapped bead in 1.5 $wt\%$ PEO solution,
	compared to corresponding best-fit model MSDs (blue solid and dashed lines) and covariances (red solid and dashed lines).
	The  experimental MSD and covariance were determined from 12650 79-step tracks.
	Errors (standard errors of the mean) are smaller than the plotted points, and therefore no error bars are plotted.
	The MSD and the covariance were fit together in two ways.
	For the fits shown as the solid lines, four fitting parameters were used, namely  $\alpha$, $D$, $\sigma^2$, and $\Delta t_E$, yielding
	 best-fit parameter values of $\alpha=0.20\pm0.13$, $D=3.3\pm2.4\times 10^{-5}~{\rm \mu m^2/s^{0.2}}$, $\sigma^2=2.8\pm2.9\times 10^{-7}~{\rm~\mu m^2}$
	 and $\Delta t_E=0.05\pm0.03~{\rm ms}$, yielding $\chi^2=110$, corresponding to $r^2$ = $9.1\times 10^{-6}~\mu$m$^2$, $\frac{\sigma^2}{r^2} = 0.031$, and RMS step size = $3.0$ nm.
	 For the fits shown as dashed lines, $\Delta t_E$ was set equal to zero, leaving three fitting parameters, $\alpha$, $D$, and $\sigma^2$. 
	 The best-fit parameter values in this case  are $\alpha=0.42$, $D=8.7\times 10^{-5}~{\rm \mu m^2/s^{0.42}}$, and $\sigma^2=2.8\times 10^{-7}~{\rm~\mu m^2}$,
	 yielding $\chi^2= 1.7\times 10^4$.}

\end{figure}

In Fig.~\ref{fig:OTfit_PEO},  we compare the measured covariance and MSD to the corresponding theoretically expected quantities (Appendix \ref{AppendixE}).
Plotted in this figure as circles and squares are the mean covariance and mean MSD, respectively, determined from 12650 79-step tracks, for the first twenty time delays.
In this case too,  we carried out global fits
of the theoretically expected forms (Eq.~\ref{BlurredMSDn}, and Eq.~\ref{eq:S0_fBM} through \ref{eq:Sn_fBM}),
first, allowing for a  non-zero $\Delta t_E$ and varying
 $\alpha$, $\Delta t_E$, $D$, and $\sigma^2$, yielding best fits with  $\chi^2=110$,  shown as the solid lines in the figure,
 and, second, setting $\Delta t_E =0$, varying  $\alpha$,  $D$, and $\sigma^2$ as fitting parameters, yielding best fits with  $\chi^2= 1.7\times 10^4$, shown as the dashed lines in the figure.
Evidently, in this example too, including motion blur provides a superior description of the experimental measurements.
Here, we also find that the best fit value of $\Delta t_E$ is close to 0.05~ms.

\subsection{Measurements and simulations of a chromosomal locus in fission yeast}
\label{sec:bio}
\subsubsection{Experimental measurements of a chromosomal locus}
\label{sec:bio_mmf1}
Next, we sought to apply the theory of Sec.~\ref{Sec:Theory} to the motion of a fluorescently-labeled genetic locus.
We examined the motion of a specific DNA locus visualized by the \textit{lacO}/GFP-lacI system in cells of the live fission yeast {\em S. pombe} (MKSP2039) \cite{Leland2014}. Specifically, a \textit{lacO} array was integrated near the \textit{mmf1} gene on chromosome II (at 3,4442,981 bp position), which is approximately in the middle between the centromeres and telomeres of the chromosome. Typically, one fluorescent focus per cell is observed. 
Fluorescence and bright field images were acquired at 30C on a DeltaVision widefield microscope (Applied Precision/GE) equipped with a temperature control chamber, a 1.4 NA,   $\times100$ objective (Olympus), solid-state-based illumination (Lumencor), and an
Evolve 512 EMCCD camera (Photometrics).
Fluorescence was excited at 488~nm and collected with emission filters passing 500-550~nm.

We analyzed 19 video microscopy movies, each containing 1000 images.
Each image was acquired for an integration time, $\Delta t_E = 10$~ms and was separated from the next image by $\Delta t = 58$~ms \cite{Bailey2020}.
A total of 157871 time steps were included in the analysis.
The position of each labeled locus was then tracked, as described in Ref.~\cite{Koo2015}, and
the resultant trajectories were partitioned into tracks of length $N=19$, 29, 39, 49, 59, 69, 79, 89, 99, 109, 159, 319, and 419 steps, as above.

\begin{figure} 
	\includegraphics[scale=0.45]{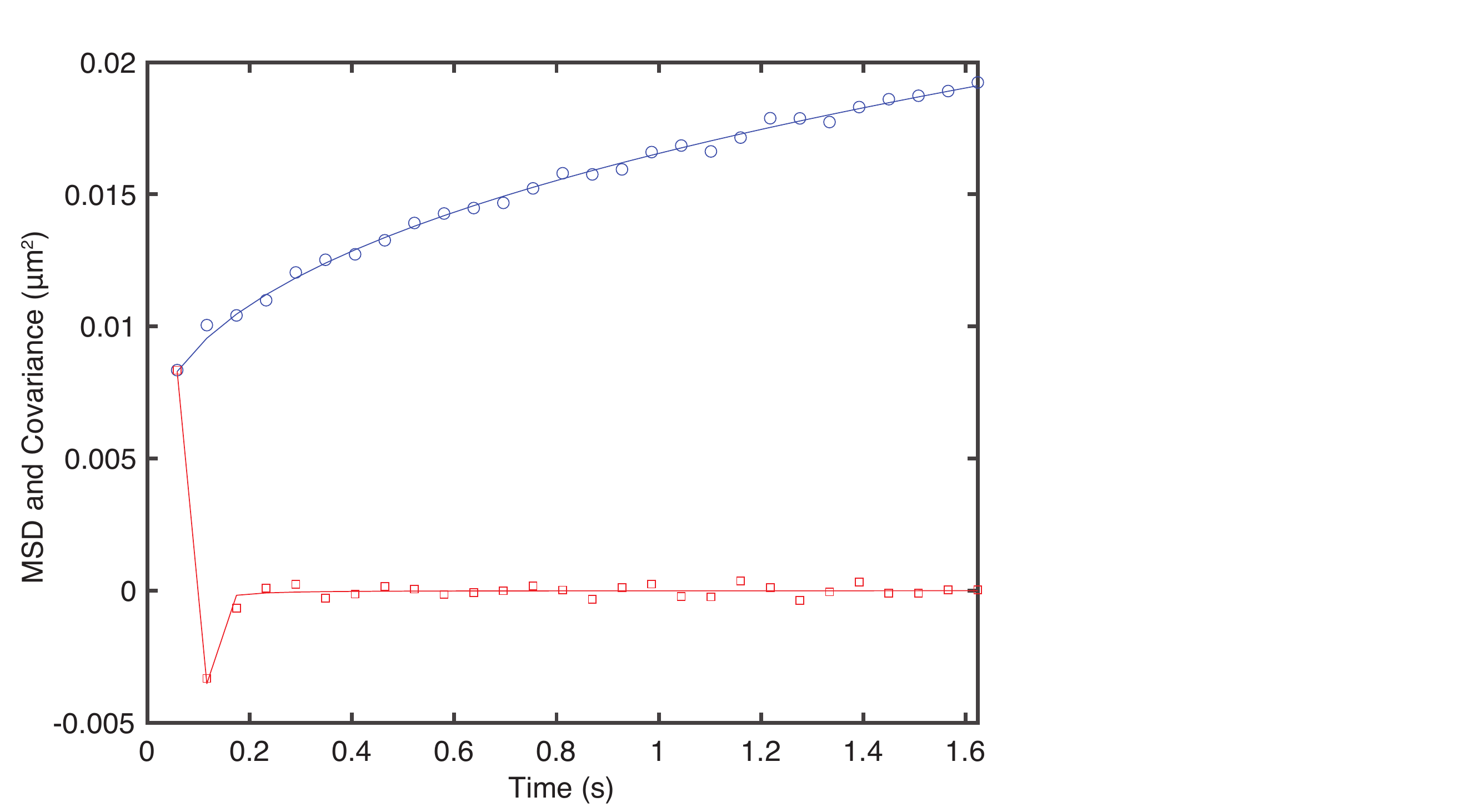}
	\caption{\label{fig:mmf1_fit} Experimental MSD (circles) and covariance (squares) versus time delay for the gene locus \textit{mmf1} in \textit{S. pombe},
		compared to the corresponding best-fit model MSD (blue solid line) and covariance (red solid line).
		The  experimental MSD and covariance were determined from 5398 29-step tracks.
		Errors (standard errors of the mean) are smaller than the plotted points, and therefore no error bars are plotted.
		The MSD and the covariance were fit together, using three fitting parameters, namely  $\alpha$, $D$, and $\sigma^2$. The resulting
		best-fit parameter values were $\alpha=0.39\pm 5\times 10^{-4}$, $D=0.0061 \pm 9\times 10^{-4}~\mu$m$^2$s$^{-1}$, and $\sigma^2=0.0027\pm 0.08~\mu$m$^2$, yielding $\chi^2=166$.}
	
\end{figure}

As also described in Ref.~\cite{Bailey2020}, the locus MSD is well-described by a model corresponding to fBm,
chartacterized by an exponent $\alpha \simeq 0.44$.
In Fig.~\ref{fig:mmf1_fit}, we compare the measured covariance (blue circles) and MSD (red squares), determined from 5398 29-step tracks to the corresponding theoretically expected quantities (Appendix \ref{AppendixE}), similar to the analysis done on the optical tweezers data in Sec.~\ref{BeadInPEO}.
We carried out global fits
of the theoretically expected forms (Eq.~\ref{BlurredMSDn} through \ref{eq:Sn_fBM}), varying
$\alpha$, $D$, and $\sigma^2$, yielding best fits with $\chi^2=166$, shown as the solid lines in the figure.
The resulting best-fit parameter values were $\alpha=0.39\pm 5\times 10^{-4}$, $D=0.0061 \pm 9\times 10^{-4}~\mu$m$^2$s$^{-1}$, and $\sigma^2=0.0027\pm 0.08~\mu$m$^2$, corresponding to $r^2$ = $0.0080~\mu$m$^2$, $\frac{\sigma^2}{r^2} = 0.34$, and RMS step size = $89.4$ nm.

\begin{figure} 
	\includegraphics[scale=0.450]{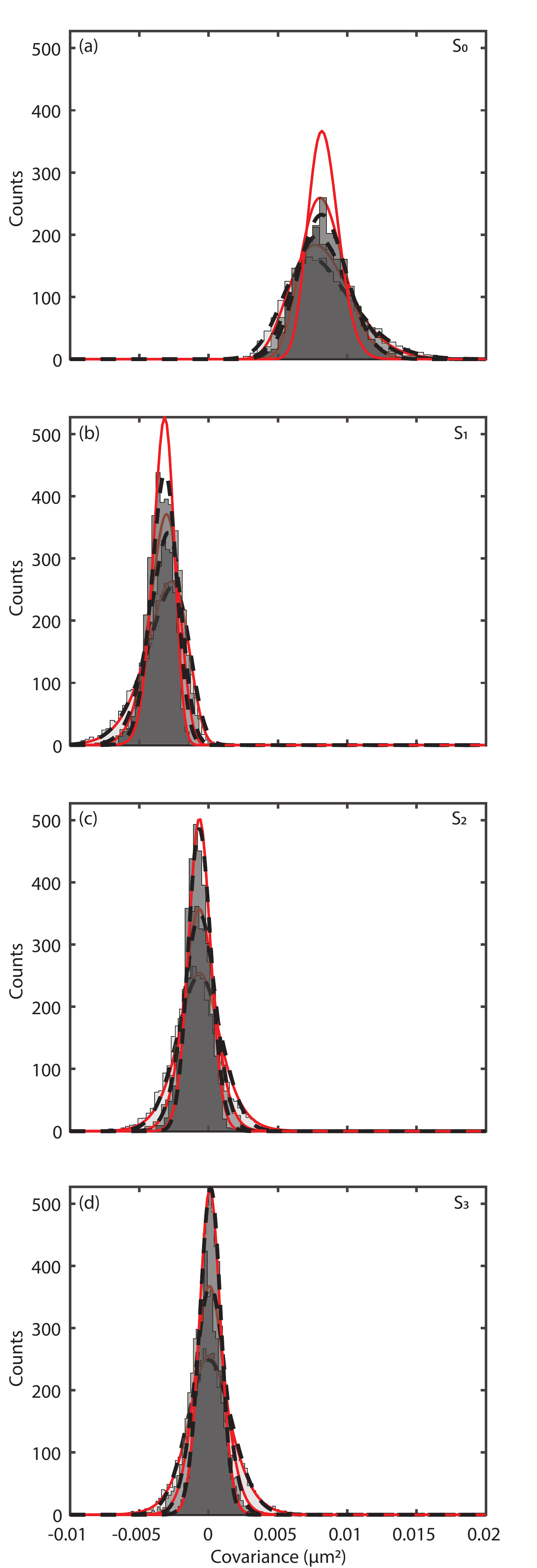}
	\caption{\label{fig:fig1_mmf1} Covariance distributions for gene locus \textit{mmf1} in \textit{S. pombe}. Covariances ${S}_0$ (a), ${S}_1$ (b), ${S}_2$ (c), and ${S}_3$ (d) for particle tracks of 19 (light gray), 39 (medium gray), and 79 (dark gray) steps are represented as histograms. Red lines correspond to the theoretical distributions. Black dashed lines correspond to the best fit of a skew normal distribution to the simulated distributions. With increasing number of time steps, the distribution narrows.  }
\end{figure}

\begin{figure}
	\includegraphics[scale=0.40]{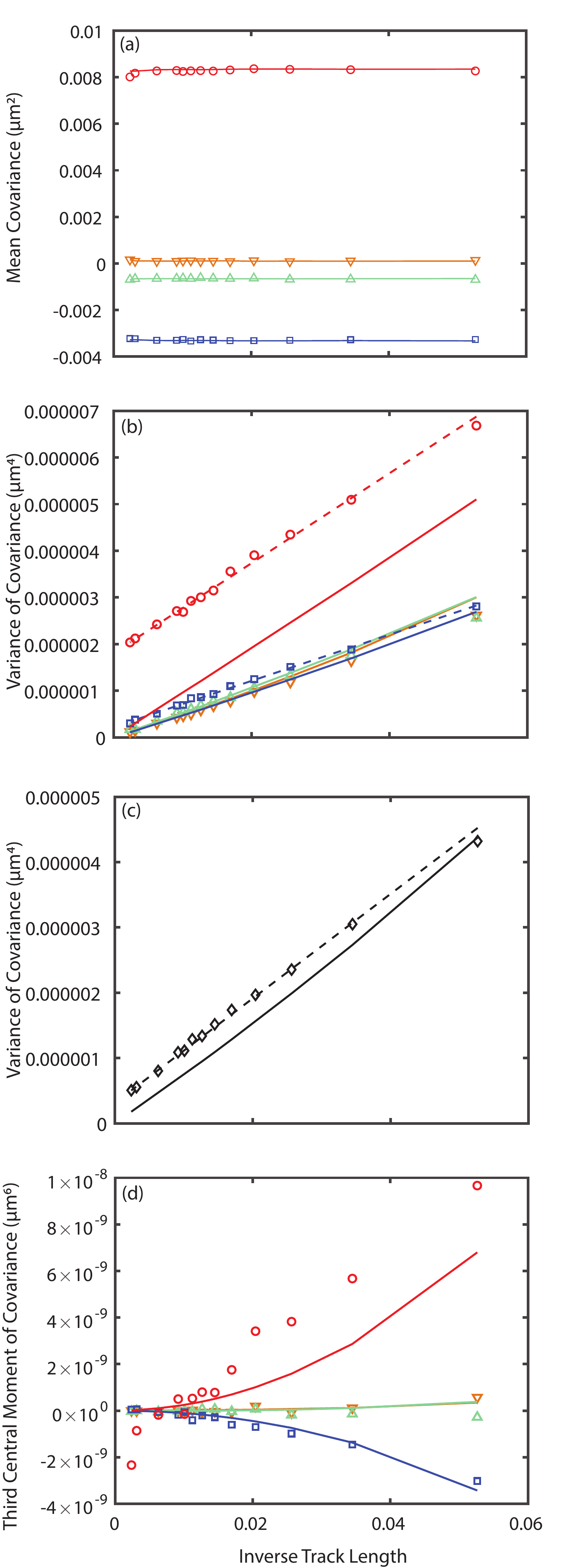}
	\caption{\label{fig:fig2_mmf1} Dependence of the covariance matrix estimates on the track length for the gene locus \textit{mmf1} in \textit{S. pombe}. (a) The mean of the covariances $S_0$ (red circle),  $S_1$ (blue square),  $S_2$ (green triangle), and  $S_3$ (orange upside-down triangle) vs. inverse track length. Theory (straight lines) is calculated by Eq.~\ref{mean}. Maximum y-value corresponds to $1.25$ in units of the mean-square step size ($8.0\times 10^{-3}$ $\mu$m$^2$). (b) Variance of the covariance vs. inverse track size of  ${S}_0$ (red circle),  ${S}_1$ (blue square),  ${S}_2$ (green triangle), and  ${S}_3$ (orange upside-down triangle). The theory is calculated by Eq.~\ref{Variance2D}. Dashed lines correspond to a fit of the experimental data to the theoretical form plus a constant. Maximum y-value corresponds to $0.11$ in units of the mean-square step size ($8.0\times 10^{-3}$ $\mu$m$^4$). (c) Variance of the covariance vs. inverse bin size of ${S}_0 + 2{S}_1$ (diamond). Calculating ${S}_0 + 2{S}_1$ removes static localization noise. Hence, the data variances are closer in value to the theory, compared the plot of ${S}_0$. Theory (solid line) is calculated using $\frac{{\rm Tr}({\bf \Sigma} ({\bf C}_0+2{\bf C}_1))^2}{4}$. Dashed lines correspond to a fit of the experimental data to the theoretical form plus a constant. Maximum y-value corresponds to $0.078$ in units of the mean-square step size ($8.0\times 10^{-3}$ $\mu$m$^4$). (d) The third central moment of the covariance ${\Sigma}_0$ (circle),  ${\Sigma}_1$ (square),  ${\Sigma}_2$ (triangle), and  ${\Sigma}_3$ (upside-down triangle) vs. inverse track length. Theory (lines) is calculated by Eq.~\ref{Skewness2D}. Maximum y-value corresponds to $0.020$ in units of the mean-square step size ($8.0\times 10^{-3}$ $\mu$m$^6$)} 
\end{figure}

The measured covariance distributions, together with the corresponding theoretical predictions, based on the mean covariance,
and the corresponding best-fits to a skew normal distribution are shown in Fig.~\ref{fig:fig1_mmf1}.
In contrast to the previous examples, the theoretical curves (red) for the $S_0$ distribution predict a significantly narrower distribution, than found experimentally (histograms).
A similar trend, albeit less pronounced, can be discerned for the $S_1$ distributions.
By contrast, the theoretical curves seem to accurately represent the experimental $S_2$ and $S_3$ distributions.
The means and variances,  determined from skew normal fits to these  distributions, are
plotted in Fig.~\ref{fig:fig2_mmf1}(a) and \ref{fig:fig2_mmf1}(b), and provide further insight.
As expected, the mean covariance is independent of track length, and the variance in track length increases with decreasing track length.
However,
in this case, the measured variance of $S_0$ is much larger than predicted, especially for long track lengths (Fig.~\ref{fig:fig2_mmf1}(b) red circles and line). 
The measured variance of $S_1$ is also noticeably larger than predicted, but to a lesser extent.
In contrast, the measured variances of $S_2$ and $S_3$ seem well-described by theory.
It appears that the variances of both $S_0$ and $S_1$ display more-or-less constant offsets above their predicted values. 
To test this idea, we fit the experimental covariances of $S_0$ and $S_1$ versus inverse track length to the theoretical form plus a constant.
The results of these fits are shown as the dashed lines in  Fig.~\ref{fig:fig2_mmf1}(b) and \ref{fig:fig2_mmf1}(c), and indeed describe the data well.
%
In general,
localization noise contributes $2 \sigma^2$ to
$\Sigma_0$ and $-\sigma^2$ to $\Sigma_1$ but does not contribute to $\Sigma_n$ for $n>1$ (Eqs.~\ref{EQ-26} and \ref{EQ-27}).
Although perhaps not the sole source of the disparity, this observation suggests that the discrepancy may originate from the localization noise in this
experimental system. This idea is further bolstered by the observation that $S_2$ and $S_3$ are well described by theory. Given that localization noise has no contribution to the mean of $\Sigma_0+2\Sigma_1$, we compared experimental and predicted variances of $\Sigma_0+2\Sigma_1$ to test whether the discrepancy with theory stems from the localization noise. 
Plotted in Fig.~\ref{fig:fig2_mmf1}(c) versus inverse track length are the experimental variance of $\Sigma_0+2\Sigma_1$, (diamonds) and the corresponding theoretical prediction (solid line).
Although a discrepancy between experiment and theory remains, the actual value of the offset is about a factor of four smaller 
than the offset observed for the variance of $S_0$ alone,  bolstering the idea that static localization noise is the culprit.

At the same time, although fluorescent foci in different cells correspond to the same genetic locus (i.e. they all correspond to the same position along the chromosome), our experimental measurements reveal that each focus may show a different fluorescence intensity and width (data not shown), which we ascribe to  different focal planes, different optical environments, {\em etc.} from cell to cell. These cell-to-cell variations in focus intensity and width surely lead to corresponding cell-to-cell variations in localization noise. Thus, our measurements themselves directly point to the possibility of cell-to-cell heterogeneity in localization noise, and thus a degree of variation in the covariance matrix $\mathbf{\Sigma}$ from cell to cell.
However, the theory presented in this paper is predicated on Eq.~\ref{multivariate}, which assumes that all samples are drawn from the same multivariate Gaussian distribution.
To investigate the possible role of localization noise heterogeneity, we carried out additional simulations,
described in Sec.~\ref{fBM-hetero}.


\subsubsection{Simulated fractional Brownian motion data with heterogeneous localization noise}
\label{fBM-hetero}

To incorporate localization noise heterogeneity, we carried out simulations
of fBm,
 for
which the localization error varies from track to track, determined via two Gaussian random variables as follows.
We consider that each track has its own unique localization noise described by a Gaussian (referred to as $G1$) with zero mean and a standard deviation determined randomly from another Gaussian distribution with the mean equal to the experimentally determined static localization noise $\left < \sigma^2 \right > = 0.0032~\mu$m$^2$, and a standard deviation equal to $\sigma_{\sigma^2}$, which is a variable parameter of the model.
Thus, each track
has its own unique localization noise.
The localization noise heterogeneity may be quantified by the value of $\frac{\sigma_{\sigma^2}}{\left < \sigma^2 \right >}$,
which we term the ``localization heterogeneity''.

\begin{figure} 
	\includegraphics[scale=0.45]{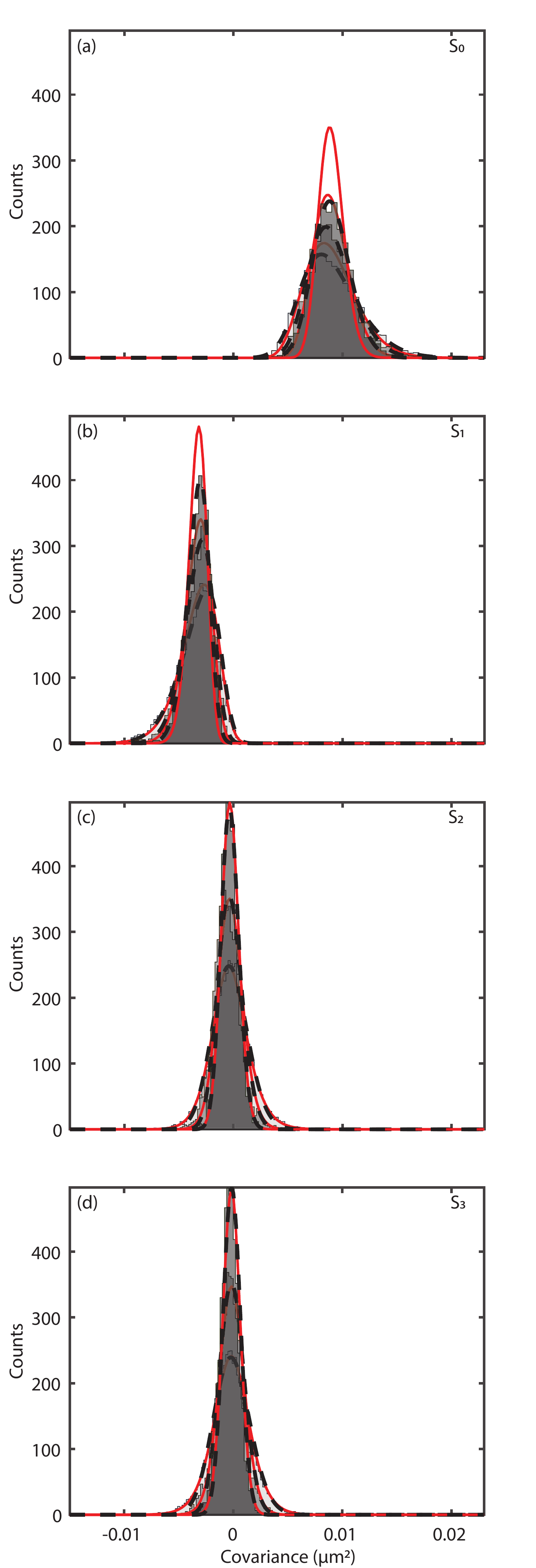}
	\caption{\label{fig:fig1_factor} Covariance distributions for simulated tracks undergoing fBm, with heterogeneous noise determined from a Gaussian distribution of widths for each track. The width of the noise distribution is determined using $homogeneity = \frac{\sigma_{\sigma^2}}{\sigma^2} =0.20$. Covariances ${S}_0$ (a), ${S}_1$ (b), ${S}_2$ (c), and ${S}_3$ (d) for particle tracks of 19 (light gray), 39 (medium gray), and 79 (dark gray) steps are represented as histograms. Red lines correspond to the theoretical distributions. Black dashed lines correspond to the best fit of a skew normal distribution to the simulated distributions. With increasing number of time steps, the distribution narrows.  }
\end{figure}

For each of several different inhomogeneities,
we simulated 1000  500-step tracks, for
$D=0.0055~\mu$m$^2$s$^{-0.44}$, $\alpha = 0.44$, $\sigma^2 = 0.0032~\mu$m$^2$, and $\Delta t = \Delta t_E = 0.058s$ ($r^2$ = $0.0063~\mu$m$^2$, $\frac{\sigma^2}{r^2} = 0.51$, and RMS step size = $79.4$ nm),
comparable to the values determined from the experiment \cite{Bailey2020}.
The resultant trajectories were partitioned as usual into tracks of length 19, 29, 39, 49,..... 319, 419, and the resultant covariance distributions fit by a skew normal function.
%
%
Fig.~\ref{fig:fig1_factor}
shows histograms of simulated $S_0$, $S_1$, $S_2$ and $S_3$ distributions, corresponding to 
$\frac{\sigma_{\sigma^2}}{\left < \sigma^2 \right >} = 0.20$.
Also shown in Fig.~\ref{fig:fig1_factor} are the corresponding theoretical  predictions (red lines)
and the skew normal best-fits (black dashed lines), which
describe the simulated distributions well in this case too.
By contrast, the theoretical curves are significantly narrower,
reminiscent of the disparity seen in Fig.~\ref{fig:fig1_mmf1} for the experimental gene locus data.
Also similar to the gene locus data, the simulated and theoretical distributions of $S_2$ and $S_3$ agree.

\begin{figure}
	\includegraphics[scale=0.4]{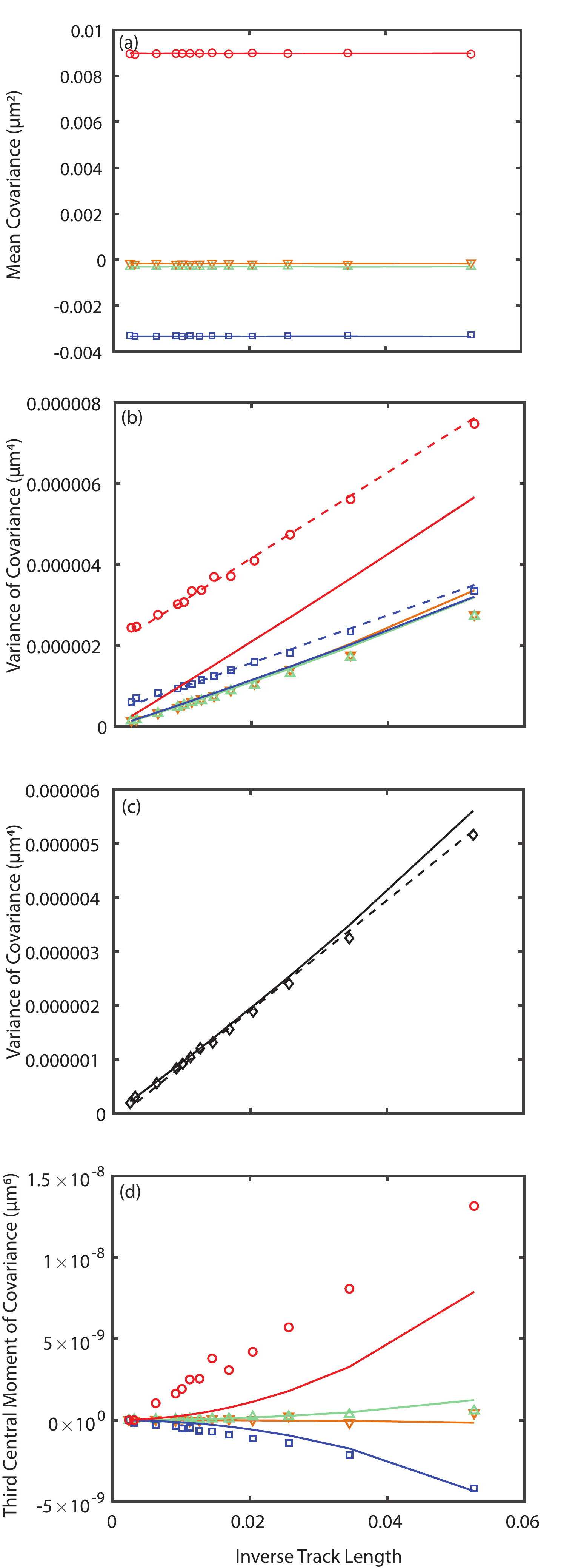}
	\caption{\label{fig:fig2_factor} Dependence of the covariance matrix estimates on the track length for simulated particles undergoing fractional Brownian motion, with noise determined from a Gaussian distribution of widths for each track. The width of the noise distribution G1 is determined using $homogeneity = \frac{\sigma_{\sigma^2}}{\sigma^2} =0.20$.
		(a) The mean of the covariances $S_0$ (red circle),  $S_1$ (blue square),  $S_2$ (green triangle), and  $S_3$ (orange upside-down triangle) {\em vs.} inverse track length. Max y-value corresponds to $1.59$ in units of the mean-square step size ($6.3\times 10^{-3}$ $\mu$m$^2$). (b)Variance of the covariance vs. inverse track size of  ${S}_0$ (red circle),  ${S}_1$ (blue square),  ${S}_2$ (green triangle), and  ${S}_3$ (orange upside-down triangle). The theory is calculated by Eq.~\ref{Variance2D}. Dotted lines represent a linear fit plus a constant to the data points. Max y-value corresponds to $0.20$ in units of the mean-square step size ($6.3\times 10^{-3}$ $\mu$m$^4$).
		(c) Variance of the covariance vs. inverse bin size of ${S}_0 + 2{S}_1$ (diamond). 
		Theory (solid line) is calculated using $\frac{{\rm Tr}({\bf \Sigma} ({\bf C}_0+2{\bf C}_1))^2}{4}$. Max y-value corresponds to $0.15$ in units of the mean-square step size ($6.3\times 10^{-3}$ $\mu$m$^4$). (d) The third central moment of the covariance ${\Sigma}_0$ (circle),  ${\Sigma}_1$ (square),  ${\Sigma}_2$ (triangle), and  ${\Sigma}_3$ (upside-down triangle) vs. inverse track length. Theory (lines) is calculated by Eq.~\ref{Skewness2D}. Max y-value corresponds to $0.06$ in units of the mean-square step size ($6.3\times 10^{-3}$ $\mu$m$^6$). } 
\end{figure}

The means, variances and 3rd moment of the covariances calculated from the best-fits for the simulated data  are shown in Fig.~\ref{fig:fig2_factor}a-d as a function inverse track length. The theory (Eq.~\ref{Variance2D}) and the theory plus offset are shown as solid and dashed lines, respectively. The means reproduce what is expected from the given simulation parameters. Note, that this is true regardless of the values of the localization heterogeneity (data not shown).  Strikingly reminiscent of the experimental results shown in Fig.~\ref{fig:fig2_mmf1}, while the variance of $S_0$ lies above the theoretically predicted values by a relatively large offset, the variance of $S_1$ also lies above that predicted theoretically, but by a relatively small offset.
The fact that our simple simulations recapitulate important features of the experimental results on gene loci  support the hypothesis that the observed theory-experiment discrepancy is the result of inhomogeneous localization noise.
To further test our model, we also determined the simulated and theoretical variance of  $\Sigma_0 + 2 \Sigma_1$ which are plotted together in Fig.~\ref{fig:fig2_factor}(c) as the diamonds and line, respectively.
In this case, however, simulation and theory match well, in contrast to the experiment-theory comparison which shows a residual discrepancy (Fig.~\ref{fig:fig2_mmf1}(c)), indicating that our simple model  does not perfectly model the experimental situation.



\begin{figure}
	\includegraphics[width=3.2in]{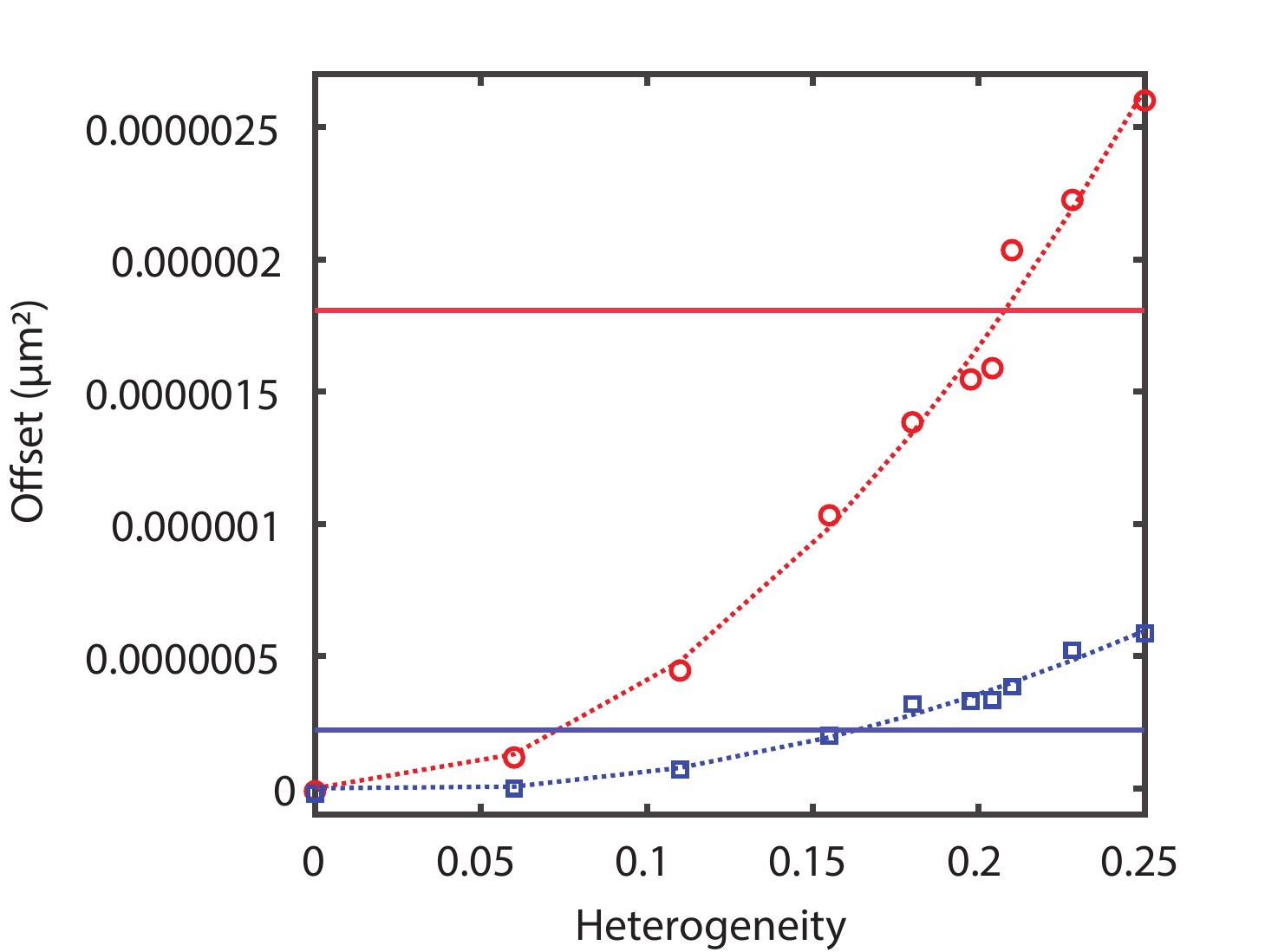}
	\caption{\label{fig:fig_offsets} Offsets between theoretical and fitted variances of simulated data with inhomogeneous localization noise for ${S}_0$ (red circle) and  ${S}_1$ (blue square), versus heterogeneity of the localization used to determine the Gaussian noise distribution. Dashed lines represent a guide to the eye for each $S_n$ offset.
	The horizontal red and blue solid lines are the  $S_0$ and $S_1$ offsets for gene locus \textit{mmf1}, respectively. 
	} 
\end{figure}

Next, we explore how localization heterogeneity levels affect the discrepancy between theory and the simulations. We vary the heterogeneity level, and, for each simulation, fit $S_0$ and $S_1$ by the theory plus offset and plot corresponding offsets vs. $\frac{\sigma_{\sigma^2}}{\sigma^2}$ (Fig.~\ref{fig:fig_offsets}). Fig.~\ref{fig:fig_offsets} plots the offsets in $S_0$ (red) and $S_1$ (blue) versus $\frac{\sigma_{\sigma^2}}{\sigma^2}$.
Unsurprisingly, small locus-to-locus variation in the localization noise results in small offsets; and ramping up the localization noise increases offsets.
For increasing inhomogeneity, {\em i.e.}
a broader initial Gaussian, G1, the offset increases.
Also shown in  Fig.~\ref{fig:fig_offsets} are the offsets for the experimental \textit{mmf1} gene locus data, where the red horizontal line is the $S_0$ offset, and the blue horizontal line is the $S_1$ offset.
The lines derived from experiment intersect the simulated curves at comparable values of the heterogeneity, namely
$\frac{\sigma_{\sigma^2}}{\sigma^2}=0.21$ for $\Sigma_0$ and $\frac{\sigma_{\sigma^2}}{\sigma^2}=0.16$ for $\Sigma_1$.
In view of the simplicity of our model, we regard this small discrepancy as satisfactory.
Distributions alternative to Gaussians yielded similar results.

\section{Conclusion}
We have developed and applied a simple method to determine whether a population of finite-length single particle tracks exhibits a single mode of diffusion. First, we derived  theoretical equations that describe the distribution of displacement covariance matrix elements for a particle in an arbitrary, but well-defined, diffusive state.
For both 1D and 2D, we determined the probability distribution of the covariance matrix elements, $S_n$, given the covariance matrix, and then calculated the first three moments of $S_n$, namely, the mean
 (trivially equal to
$\Sigma_n$), variance, and third central moment. 
To test these theoretical results, we first simulated tracks undergoing simple 2D diffusion,  uncovering excellent agreement between theoretical and simulated covariance matrix element distributions.
We further showed that least-mean-squares fits of a skew normal function to the simulated distributions were able to
accurately capture the shape of the simulated distributions.  Best-fit parameters from
skew normal fits, carried out for different track lengths also matched well to the theoretical expressions for the variance and third central moment versus  track length.
In addition to simulated tracks, we also followed this procedure for experimental data,  first for an optically-trapped bead in water and then an optically-trapped bead in a viscoelastic high polymer solution. In these cases also,
we found that the experimental covariance distributions are well described by the theoretical covariance distributions,
using the measured mean covariances as the sole input. In these cases too, skew normal best fits yielded
parameters that well described the mean, variance, and third central moment of the experimental covariance
distributions versus inverse track length.  These collected results give us confidence that the theory, presented
in Sec. 2, is correct.

Finally, we analyzed the covariance distributions of the \textit{mmf1} gene locus in \textit{S. pombe}. 
Here, the analysis revealed that the experimental $S_0$ and $S_1$ distributions are significantly wider than predicted on the basis of the mean covariance matrix elements, initially suggesting the presence of more than one diffusive state. However, the fact that the discrepancy is confined only to  $S_0$ and $S_1$ led us to consider the hypothesis that the discrepancy originates with locus-to-locus heterogeneity in the static  localization noise. To test this idea,
we simulated fBm tracks with static localization noise of varying degrees of heterogeneity. Analyses of these tracks recreated a similar disparity between the theoretical and measured variance of $S_0$ and $S_1$ versus inverse track length. We explored varying the degree of the heterogeneity in the context of simulations
in an attempt to more closely recapitulate the discrepancy seen in the gene locus data. While  the heterogeneous noise we added largely accounts
for the offsets we saw between the measured and theoretical variances of the covariance, it did not fully account for this behavior. Further investigation is needed to completely understand the factors that contribute to this discrepancy.
In summary, we found that, for a single mode of diffusion, covariance distributions can be accurately predicted using the presented theory and mean covariances. Comparison of the predicted distributions with the experimentally measured ones enabled us to answer whether particles display a single or multiple diffusive states. Hence, this method is a simple yet powerful tool to determine whether a biological system exhibits a single diffusive state.  

\begin{acknowledgments}
 This research was supported by 
 NSF CMMI 1634988 and 
 NSF EFRI CEE award EFMA-1830904.
 M. L. P. B. was supported by NIH T32EB019941 and the NSF GRFP.
%
 %
%
\end{acknowledgments}

\appendix
\begin{widetext}

\section{Calculation of the central moments generalized to k-dimensional space}
\label{ndspace}
The probability distribution of $S_n$ in k-dimensional space is
\begin{eqnarray}
P(S_n | \mathbf{\Sigma} ) &=
  \int^\infty_{-\infty} \frac{d \omega}{2 \pi }
\frac{1}{   |{\bf I}+  \frac{i}{k} \omega \mathbf{\Sigma} {\bf C}_n |^{k/2} }
e^{i \omega S_n}.
\end{eqnarray}

Expand $\lvert {\bf I}+\frac{i}{k}\omega{\mathbf \Sigma} {\bf C}_n\rvert $ to the third order,
\begin{eqnarray}
\lvert {\bf I}+\frac{i}{k}\omega{\mathbf \Sigma} {\bf C}_n\rvert &=&
\lvert{\mathbf  \Sigma} {\bf C}_n \rvert \lvert({\mathbf \Sigma} {\bf C}_n)^{-1}+\frac{i}{k}\omega {\bf I} \rvert \nonumber \\
&=&
\Lambda_1 \Lambda_2 \Lambda_3 ... \Lambda_N \left( \Lambda_1^{-1}+\frac{i}{k}\omega\right) ... \left(\Lambda_N^{-1}+\frac{i}{k}\omega\right) \nonumber \\
&=&
1+\left(\Lambda_1+ \Lambda_2+ ... +\Lambda_N\right) (\frac{i}{k}\omega) \nonumber \\
&& +\left(\Lambda_1\Lambda_2+\Lambda_1 \Lambda_3+...+\Lambda_1 \Lambda_N+...+\Lambda_{N-1} \Lambda_N\right)\left(\frac{i}{k}\omega\right)^2   \nonumber \\
&&+\left(\Lambda_1\Lambda_2\Lambda_3 +\Lambda_1\Lambda_2\Lambda_4+...+\Lambda_{N-2}\Lambda_{N-1} \Lambda_N\right)\left(\frac{i}{k}\omega\right)^3 +O(\omega^4)\nonumber \\
&=&
1+Tr({\mathbf \Sigma} {\bf C}_n)(\frac{i}{2}\omega)+ 
\frac{(Tr {\mathbf \Sigma} {\bf C}_n)^2-Tr({\mathbf \Sigma} {\bf C}_n)^2}{2}\left(\frac{i}{k}\omega\right)^2  \nonumber\\
&&+ \frac{(Tr {\mathbf \Sigma} {\bf C}_n)^3-3Tr({\mathbf \Sigma} {\bf C}_n)^2Tr({\mathbf \Sigma} {\bf C}_n)+2Tr({\mathbf \Sigma} {\bf C}_n)^3}{6}\left(\frac{i}{k}\omega\right)^3  
+O(\omega^4).\nonumber\\
\end{eqnarray}

It follows that
\begin{eqnarray}
\frac{d}{d\omega} \lvert {\bf I}+\frac{i}{k}\omega {\mathbf \Sigma} {\bf C}_n\rvert\bigg|_{\omega=0}  =\frac{i}{k} Tr({\mathbf \Sigma} {\bf C}_n),
\label{FirstDerive32}
\end{eqnarray}
\begin{eqnarray}
\frac{d^2}{d\omega^2} \lvert {\bf I}+\frac{i}{k}\omega {\mathbf \Sigma} {\bf C}_n \rvert \bigg|_{\omega=0} = (\frac{i}{k})^2 \Big( (Tr {\mathbf \Sigma} {\bf C}_n)^2-Tr({\mathbf \Sigma} {\bf C}_n)^2\Big),
\label{SecondDerive32}
\end{eqnarray}
and
\begin{eqnarray}
\frac{d^3}{d\omega^3} \lvert {\bf I}+\frac{i}{k}\omega {\mathbf \Sigma} {\bf C}_n\rvert\bigg|_{\omega=0}  =(\frac{i}{k})^3\Big( (Tr {\mathbf \Sigma} {\bf C}_n)^3-3Tr({\mathbf \Sigma} {\bf C}_n)^2Tr({\mathbf \Sigma} {\bf C}_n)+2Tr({\mathbf \Sigma} {\bf C}_n)^3\Big).\nonumber\\
\label{ThirdDerive32}
\end{eqnarray}

\begin{eqnarray}
\left < S_n \right > &=& \int_{-\infty}^{\infty} d S_n  P(S_n \vert {\mathbf \Sigma}) S_n \nonumber \\
&= &
\int_{-\infty}^{\infty} \frac{d\omega}{2\pi} \frac{1}{\lvert {\bf I}+\frac{i}{k}\omega {\mathbf \Sigma} {\bf C}_n\rvert^{k/2}} \int_{-\infty}^{\infty} d S_n  S_n  e^{iS_n\omega}  \nonumber \\
&=&
\int_{-\infty}^{\infty} \frac{d\omega}{2\pi} \int_{-\infty}^{\infty} dS_n \frac{1}{\lvert {\bf I}+\frac{i}{k}\omega {\mathbf \Sigma} {\bf C}_n\rvert^{k/2}} \bigg( \frac{d}{id\omega} e^{iS_n\omega} \bigg) \nonumber \\
&=&
-\int_{-\infty}^{\infty} d\omega \int_{-\infty}^{\infty} \frac{1}{2\pi} dS_n e^{iS_n\omega} \frac{d}{id\omega} \frac{1}{\lvert{\bf  I}+\frac{i}{k}\omega {\mathbf \Sigma} {\bf C}_n \rvert^{k/2}}  \nonumber \\
&=& -\frac{d}{id\omega} \frac{1}{\lvert{\bf  I}+\frac{i}{k}\omega {\mathbf \Sigma} {\bf C}_n \rvert^{k/2}}  \bigg|_{\omega=0} \nonumber\\
&=&
\frac{k}{2}\frac{1}{\lvert{\bf  I}+\frac{i}{k}\omega {\mathbf \Sigma} {\bf C}_n \rvert^{k/2+1}} \frac{d}{id\omega} \lvert{\bf  I}+\frac{i}{k}\omega {\mathbf \Sigma} {\bf C}_n \rvert\bigg|_{\omega=0} \nonumber\\
&=& \frac{1}{2} Tr({\mathbf \Sigma} {\bf C}_n).
\label{EQSnk}
\end{eqnarray}

\begin{eqnarray}
\left < S_n^2 \right > &=& \int_{-\infty}^{\infty} d S_n  P(S_n \vert {\mathbf \Sigma}) S_n^2 \nonumber \\
&= &
\int_{-\infty}^{\infty} \frac{d\omega}{2\pi} \frac{1}{\lvert {\bf I}+\frac{i}{k}\omega {\mathbf \Sigma} {\bf C}_n\rvert^{k/2}} \int_{-\infty}^{\infty} d S_n^2  S_n  e^{iS_n\omega}  \nonumber \\
&=&
\int_{-\infty}^{\infty} \frac{d\omega}{2\pi} \int_{-\infty}^{\infty} dS_n \frac{1}{\lvert {\bf I}+\frac{i}{k}\omega {\mathbf \Sigma} {\bf C}_n\rvert^{k/2}} \bigg( \frac{d^2}{i^2d\omega^2} e^{iS_n\omega} \bigg) \nonumber \\
&=&
\int_{-\infty}^{\infty} d\omega \int_{-\infty}^{\infty} \frac{1}{2\pi} dS_n e^{iS_n\omega} \frac{d^2}{i^2d\omega^2} \frac{1}{\lvert{\bf  I}+\frac{i}{k}\omega {\mathbf \Sigma} {\bf C}_n \rvert^{k/2}}  \nonumber \\
&=& \frac{d^2}{i^2d\omega^2} \frac{1}{\lvert{\bf  I}+\frac{i}{k}\omega {\mathbf \Sigma} {\bf C}_n \rvert^{k/2}}  \bigg|_{\omega=0} \nonumber\\
&=& \frac{d}{i^2d\omega}
\bigg[-\frac{k}{2}\frac{1}{\lvert{\bf  I}+\frac{i}{k}\omega {\mathbf \Sigma} {\bf C}_n \rvert^{k/2+1}} \frac{d}{d\omega} \lvert{\bf  I}+\frac{i}{k}\omega {\mathbf \Sigma} {\bf C}_n \rvert\bigg]\bigg|_{\omega=0} \nonumber\\
&=&  \frac{1}{i^2}\bigg[\frac{k}{2}(\frac{k}{2}+1)\frac{1}{\lvert{\bf  I}+\frac{i}{k}\omega {\mathbf \Sigma} {\bf C}_n \rvert^{k/2+2}} \bigg(\frac{d}{d\omega} \lvert{\bf  I}+\frac{i}{k}\omega {\mathbf \Sigma} {\bf C}_n \rvert\bigg)^2-\frac{k}{2}\frac{1}{\lvert{\bf  I}+\frac{i}{k}\omega {\mathbf \Sigma} {\bf C}_n \rvert^{k/2+1}} \frac{d^2}{d\omega^2} \lvert{\bf  I}+\frac{i}{k}\omega {\mathbf \Sigma} {\bf C}_n \rvert\bigg]\bigg|_{\omega=0} \nonumber\\
&=&
\frac{k}{2}(\frac{k}{2}+1)\frac{1}{k^2}(Tr{\mathbf \Sigma} {\bf C}_n)^2 -\frac{k}{2}\frac{1}{k^2}\Big[ (Tr {\mathbf \Sigma} {\bf C}_n)^2-Tr({\mathbf \Sigma} {\bf C}_n)^2\Big]\nonumber\\
&=&
\frac{1}{4}(Tr{\mathbf \Sigma} {\bf C}_n)^2 +\frac{1}{2k}Tr({\mathbf \Sigma} {\bf C}_n)^2.
\label{EQSn2k}
\end{eqnarray}

\begin{eqnarray}
\left < S_n^3 \right > &=& \int_{-\infty}^{\infty} d S_n  P(S_n \vert {\mathbf \Sigma}) S_n^3 \nonumber \\
&= &
\int_{-\infty}^{\infty} \frac{d\omega}{2\pi} \frac{1}{\lvert {\bf I}+\frac{i}{k}\omega {\mathbf \Sigma} {\bf C}_n\rvert^{k/2}} \int_{-\infty}^{\infty} d S_n^3  S_n  e^{iS_n\omega}  \nonumber \\
&=&
\int_{-\infty}^{\infty} \frac{d\omega}{2\pi} \int_{-\infty}^{\infty} dS_n \frac{1}{\lvert {\bf I}+\frac{i}{k}\omega {\mathbf \Sigma} {\bf C}_n\rvert^{k/2}} \bigg( \frac{d^3}{i^3d\omega^3} e^{iS_n\omega} \bigg) \nonumber \\
&=&-
\int_{-\infty}^{\infty} d\omega \int_{-\infty}^{\infty} \frac{1}{2\pi} dS_n e^{iS_n\omega} \frac{d^3}{i^3d\omega^3} \frac{1}{\lvert{\bf  I}+\frac{i}{k}\omega {\mathbf \Sigma} {\bf C}_n \rvert^{k/2}}  \nonumber \\
&=& -\frac{d^3}{i^3d\omega^3} \frac{1}{\lvert{\bf  I}+\frac{i}{k}\omega {\mathbf \Sigma} {\bf C}_n \rvert^{k/2}}  \bigg|_{\omega=0} \nonumber\\
&=& -\frac{d^2}{i^3d\omega^2}
\bigg[-\frac{k}{2}\frac{1}{\lvert{\bf  I}+\frac{i}{k}\omega {\mathbf \Sigma} {\bf C}_n \rvert^{k/2+1}} \frac{d}{d\omega} \lvert{\bf  I}+\frac{i}{k}\omega {\mathbf \Sigma} {\bf C}_n \rvert\bigg]\bigg|_{\omega=0} \nonumber\\
&=& - \frac{d}{i^3d\omega}\bigg[\frac{k}{2}(\frac{k}{2}+1)\frac{1}{\lvert{\bf  I}+\frac{i}{k}\omega {\mathbf \Sigma} {\bf C}_n \rvert^{k/2+2}} \bigg(\frac{d}{d\omega} \lvert{\bf  I}+\frac{i}{k}\omega {\mathbf \Sigma} {\bf C}_n \rvert\bigg)^2-\frac{k}{2}\frac{1}{\lvert{\bf  I}+\frac{i}{k}\omega {\mathbf \Sigma} {\bf C}_n \rvert^{k/2+1}} \frac{d^2}{d\omega^2} \lvert{\bf  I}+\frac{i}{k}\omega {\mathbf \Sigma} {\bf C}_n \rvert\bigg]\bigg|_{\omega=0} \nonumber\\
&=&
 - \frac{1}{i^3}\bigg[-\frac{k}{2}(\frac{k}{2}+1)(\frac{k}{2}+2)\frac{1}{\lvert{\bf  I}+\frac{i}{k}\omega {\mathbf \Sigma} {\bf C}_n \rvert^{k/2+3}} \bigg(\frac{d}{d\omega} \lvert{\bf  I}+\frac{i}{k}\omega {\mathbf \Sigma} {\bf C}_n \rvert\bigg)^3\nonumber\\
&& +3\frac{k}{2}(\frac{k}{2}+1)\frac{1}{\lvert{\bf  I}+\frac{i}{k}\omega {\mathbf \Sigma} {\bf C}_n \rvert^{k/2+2}} \frac{d}{d\omega} \lvert{\bf  I}+\frac{i}{k}\omega {\mathbf \Sigma} {\bf C}_n \rvert\frac{d^2}{d\omega^2} \lvert{\bf  I}+\frac{i}{k}\omega {\mathbf \Sigma} {\bf C}_n \rvert\nonumber\\
&&-\frac{k}{2}\frac{1}{\lvert{\bf  I}+\frac{i}{k}\omega {\mathbf \Sigma} {\bf C}_n \rvert^{k/2+1}} \frac{d^3}{d\omega^3} \lvert{\bf  I}+\frac{i}{k}\omega {\mathbf \Sigma} {\bf C}_n \rvert\bigg]\bigg|_{\omega=0} \nonumber\\
&=&
\frac{k}{2}(\frac{k}{2}+1)(\frac{k}{2}+2)\frac{1}{k^3}(Tr{\mathbf \Sigma} {\bf C}_n)^3-3\frac{k}{2}(\frac{k}{2}+1)\frac{1}{k}Tr({\mathbf \Sigma} {\bf C}_n)\frac{1}{k^2}\big[(Tr {\mathbf \Sigma} {\bf C}_n)^2-Tr({\mathbf \Sigma} {\bf C}_n)^2\big]\nonumber\\
&&+\frac{k}{2}\frac{1}{k^3}\big[(Tr {\mathbf \Sigma} {\bf C}_n)^3-3Tr({\mathbf \Sigma} {\bf C}_n)^2Tr({\mathbf \Sigma} {\bf C}_n)+2Tr({\mathbf \Sigma} {\bf C}_n)^3\big]\nonumber\\
&=&
\frac{1}{8}(Tr{\mathbf \Sigma} {\bf C}_n)^3+\frac{3}{4k}Tr({\mathbf \Sigma} {\bf C}_n)Tr({\mathbf \Sigma} {\bf C}_n)^2+\frac{1}{k^2}Tr({\mathbf \Sigma} {\bf C}_n)^3.
\label{EQSn3k}
\end{eqnarray}

Using Eq.~\ref{EQSnk},~\ref{EQSn2k} and \ref{EQSn3k}, we obtain the first central moment
\begin{eqnarray}
\mu=\left < S_n \right > = \frac{1}{2} Tr({\mathbf \Sigma} {\bf C}_n),
\end{eqnarray}
the second central moment
\begin{eqnarray}
\sigma^2_{S_n} &=& \left < S_n^2 \right > - \left < S_n \right > ^2 \nonumber \\
&=&
\frac{1}{4} (Tr{\mathbf \Sigma} {\bf C}_n)^2 +  \frac{1}{2k}Tr({\mathbf \Sigma} {\bf C}_n)^2 - \left(\frac{1}{2}Tr{\mathbf \Sigma} {\bf C}_n\right)^2 \nonumber \\
&=&
\frac{1}{2k}Tr({\mathbf \Sigma} {\bf C}_n)^2,
\end{eqnarray}
and the third central moment
\begin{eqnarray}
\mu_3&=& E[(S_n-\left <S_n\right>)^3]\nonumber\\
&=& \left <S_n^3\right>-3\left <S_n\right>\sigma^2_{S_n}-\left <S_n\right>^3\nonumber\\
&=& \frac{1}{8} (Tr{\mathbf \Sigma} {\bf C}_n)^3 +\frac{3}{4k} Tr({\mathbf \Sigma} {\bf C}_n)Tr({\mathbf \Sigma} {\bf C}_n)^2 +\frac{1}{k^2}Tr({\mathbf \Sigma} {\bf C}_n)^3\nonumber\\
&&-3\frac{Tr({\mathbf \Sigma} {\bf C}_n)}{2}\frac{Tr({\mathbf \Sigma} {\bf C}_n)^2}{2k}
-\bigg[\frac{Tr({\mathbf \Sigma} {\bf C}_n)}{2}\bigg]^3\nonumber\\
&=&\frac{1}{k^2}Tr({\mathbf \Sigma} {\bf C}_n)^3.
\end{eqnarray}

The skewness (3rd standardized moment) $\widetilde{\mu_3}$ is
\begin{eqnarray}
\widetilde{\mu_3}&=& \frac{\mu_3}{\sigma^3_{S_n}}=\frac{Tr({\mathbf \Sigma} {\bf C}_n)^3}{k^2[\frac{Tr({\mathbf \Sigma} {\bf C}_n)^2}{2k}]^{\frac{3}{2}}}.
\label{skewnessa12}
\end{eqnarray}

\end{widetext}

\section{}
\label{AppendixC}
Because all tridiagonal Toeplitz matrices possess the same eigenvectors,
the matrices, $\mathbf \Sigma$,  ${\bf C}_1$, $\mathbf{I}+\frac{i}{2} \omega \mathbf{\Sigma}\bf{C}_0$,  $\mathbf{I}+\frac{i}{2} \omega \mathbf{\Sigma}\bf{C}_1$ and
$\mathbf{I}+\frac{i}{8 \Delta t} \omega ( \mathbf{\Sigma}\bf{C}_0+ {\rm 2} \mathbf{\Sigma}\bf{C}_1)$, are all diagonalized by the same orthogonal transformation, given by
$[{\bf U}]_{j k} = \sqrt{\frac{2}{N}} \sin \frac {\pi  j k }{N+1}$.
For $\mathbf \Sigma$ and ${\bf C}_1$,  the eigenvalues are
\begin{equation}
\Lambda_k=\Sigma_0 +2 \Sigma_1 \cos \frac{ k \pi}{N+1}
\end{equation}
and
\begin{equation}
\lambda_{1 k}=\frac{2}{N-1}\cos \frac{ k \pi}{N+1},
\end{equation}
respectively,
where $k = 1$, $2$, .. $N$. 
It follows that
\begin{equation}
p(S_0|{\mathbf \Sigma})= \int^\infty_{-\infty}
 \frac{d \omega}{2 \pi}
\frac{e^{i \omega S_0 }}{
\prod_{k=1}^N 
{
1+  i \omega
\left (
\Sigma_0 +2 \Sigma_1 \cos 
\frac{
k \pi
}{
N+1
} 
 \right )
\frac{1}{N}
}
},
\label{EQ-31a}
\end{equation}
\begin{equation}
        \begin{array}{ll}
        p(S_1|{\mathbf \Sigma})
&= \int^\infty_{-\infty}
 \frac{d \omega}{2 \pi}
\frac{e^{i \omega S_1 }}{
\prod_{k=1}^N 
{
1+  i \omega
\left (
\Sigma_0 +2 \Sigma_1 \cos 
\frac{
k \pi
}{
N+1
} 
 \right )
  \frac{1}{N-1}\cos\frac{k \pi}{N+1}
}
}
,
\label{EQ-32}
\end{array}
\end{equation}
and
\begin{widetext}
\begin{eqnarray}
P(D | \mathbf{\Sigma} ) 
&= \int^\infty_{-\infty}
\frac{d \omega}{2 \pi}
\frac{e^{i \omega D}}{\
	{
		\prod_{k=1}^N 
		\left [ 1+ \frac{ i \omega}{ 4\Delta t} 
		\left (
		\Sigma_0 +2 \Sigma_1 \cos 
		\frac{
			k \pi
		}{
			N+1
		} 
		\right )
		\left ( \frac{1}{N}+ \frac{2\cos\frac{k \pi}{N+1}}{N-1} \right )
		\right ]
	}
}
,
\label{EQ-33}
\end{eqnarray}
\end{widetext}
which is the 2D version of Eq.~A11 of Ref.~\cite{Vestergaard2014}.

\section{MSD and covariance matrix elements for a particle in a harmonic potential}
\label{AppendixD}
The usual approach for analyzing optical tweezers data is to calculate the power spectrum of the displacement fluctuations
away from the potential minimum, which is equivalent to consideration of the mean-square displacement (MSD) versus time.
To determine what new information can be gleaned by considering the covariance matrix elements of optical tweezers data,
in this section, we calculate both the MSD and the covariance matrix elements for a particle in a viscous fluid, subject to a harmonic potential. Some of these results can be found in \cite{Norrelykke2011}.

Our starting point is the theoretical result for the displacement correlation function:
\begin{equation}
\left < x(t) x(s) \right > = \frac{k_B T}{\kappa} e^{-|t-s|/\tau},
\end{equation}
where $\kappa$ is the trap stiffness, $\tau$ is the characteristic time of the trap, $k_B$ is Boltzmann's constant and $T$ is the absolute temperature. 
Thus,  for times $t_0$ and $t_n$, separated by a time $n \Delta t$, we have
\begin{equation}
\left < x(t_n) x(t_0) \right > =
\left < x_n x_0 \right > =
\frac{k_B T}{\kappa} e^{-(t_n-t_0)/\tau} = \frac{k_B T}{\kappa} e^{-n \Delta t/\tau}.
\end{equation}
It follows that the MSD, $M_n$,  after $n$ time steps, each of $\Delta t$, is
\begin{widetext}
\begin{equation}
M_n=\left < ( x_n - x_0 )^2 \right >
=\left <x_n^2 \right > + \left <x_0^2 \right > -2\left <  x_n x_0 \right >
= \frac{2 k_B T}{\kappa} (1 - e^{-n \Delta t/\tau} ),
\label{EQ-37}
\end{equation}
while the covariance matrix elements are
\begin{equation}
\Sigma_0=M_0=\left < ( x_1 - x_0 )^2 \right >
=\left <x_1^2 \right > + \left <x_0^2 \right > -2\left <  x_1 x_0 \right >
= \frac{2 k_B T}{\kappa} (1 - e^{-\Delta t/\tau} ),
\label{EQ-38}
\end{equation}
\begin{equation}
\Sigma_1= \left < (x_2-x_1)( x_1 - x_0 ) \right >
=\left <x_2 x_1 \right > - \left <x_1^2 \right > +\left <  x_1 x_0 \right >- \left < x_2 x_0 \right >
= - \frac{k_B T}{\kappa} (1 - e^{-\Delta t/\tau})^2,  
\label{EQ-39}
\end{equation}
and
\begin{equation}
\Sigma_n=
 \left < (x_{n+1}-x_n)( x_1 - x_0 ) \right >
=\left <x_{n+1} x_1 \right > - \left <x_n x_1 \right > +\left <  x_{n} x_0 \right > - \left < x_{n+1} x_0 \right >
= - \frac{k_B T}{\kappa}  e^{-(n-1)\Delta t/\tau} (1 - e^{-\Delta t/\tau})^2.  
\label{EQ-40}
\end{equation}
\end{widetext}

However,  data acquisition actually occurs for a time period
 $\Delta t_E$, while the spacing between successive acquisitions  is $\Delta t$.
  Thus, rather than measuring an instantaneous particle position, in general, experiments
 measure a motion-blurred position, averaged over the duration of data acquisition,
{\em i.e.}, acquisition period $n$ measures
 \begin{equation}
\frac{1}{\Delta t_E}  \int_0^{\Delta t_E} dt ~x(n\Delta t + t),
\end{equation}
where $x(n \Delta t +t)$ is the particle position at time $n\Delta t + t$.
%
%
To incorporate motion blur into our calculations of the MSD and the covariance matrix elements,
we must replace $\left < x_0^2 \right >$, {\em etc.} in Eqs.~\ref{EQ-37} through \ref{EQ-40} by the appropriate time-averaged quantities.
Using
\begin{widetext}
\begin{equation}
\left < x_0^2 \right > = \frac{1}{{\Delta t_E}^2} \int_0^{\Delta t_E} dt \int_0^{\Delta t_E} ds ~ e^{-|t-s|/\tau}
= \frac{2}{{\Delta t_E}^2} \int_0^{\Delta t_E} dt \int_0^t ds ~ e^{-(t-s)/\tau}
=  \frac{2 (e^{-{\Delta t_E}/\tau}-1 + \frac{{\Delta t_E}}{\tau})}{{\Delta t_E}^2/\tau^2},
\end{equation}
and 
\begin{equation}
\left < x_m x_0 \right > = \frac{1}{{\Delta t_E}^2} \int_0^{\Delta t_E} dt \int_0^{\Delta t_E} ds ~ e^{-(t_m+t-t_0-s)/\tau}
= \frac{2 ( \cosh{{\Delta t_E}/\tau} -1 )}{{\Delta t_E}^2/\tau^2} e^{-(t_m-t_0)/\tau}.
\end{equation}
and incorporating static localization noise,
 the motion-blurred MSD becomes:
\begin{equation}
M_n
= \frac{4 k_B T}{\kappa} \left ( \frac{e^{-{\Delta t_E}/\tau}-1 + \frac{{\Delta t_E}}{\tau}}{{\Delta t_E}^2/\tau^2}
 - \frac{  \cosh{{\Delta t_E}/\tau} -1 }{{\Delta t_E}^2/\tau^2}e^{-n\Delta t/\tau} \right ) + 2 \sigma^2.
 \label{TweezersMSD}
\end{equation}
Similarly, the covariance matrix elements become:
\begin{equation}
\Sigma_0
= \frac{4 k_B T}{\kappa} \left ( \frac{e^{-{\Delta t_E}/\tau}-1 + \frac{{\Delta t_E}}{\tau}}{{\Delta t_E}^2/\tau^2}
 - \frac{  \cosh{{\Delta t_E}/\tau} -1 }{{\Delta t_E}^2/\tau^2}e^{-\Delta t/\tau} \right ) + 2 \sigma^2,
\label{TweezersSigma0}
\end{equation}
\begin{equation}
\Sigma_1
= - \frac{2 k_B T}{\kappa} \left ( \frac{ e^{-{\Delta t_E}/\tau}-1 + \frac{{\Delta t_E}}{\tau}}{{\Delta t_E}^2/\tau^2} +\frac{  \cosh{{\Delta t_E}/\tau} -1 }{{\Delta t_E}^2/\tau^2} e^{-\Delta t/\tau}( -2 +e^{- \Delta t/\tau}) \right ) - \sigma^2,  
\label{TweezersSigma1}
\end{equation}
and
\begin{equation}
\Sigma_n
= - \frac{2 k_B T}{\kappa}  \frac{ \cosh{{\Delta t_E}/\tau} -1}{{\Delta t_E}^2/\tau^2} e^{-(n-1)\Delta t/\tau} (1 - e^{-\Delta t/\tau})^2,  
\label{TweezersSigman}
\end{equation}
for $n>1$.
\end{widetext}
In the limit that
 $\Delta t  \ll \tau$  and ${\Delta t_E} \ll \tau$,
using $\frac{k_B T}{\kappa \tau} = D$, where $D$ is the particle's diffusion coefficient,
Eqs.~\ref{TweezersSigma0}, \ref{TweezersSigma1}, and \ref{TweezersSigman}
reproduce the corresponding results for free diffusion with motion blur and static localization noise (Eqs.~\ref{EQ-26} and \ref{EQ-27}), as expected.
Generally, power-spectrum-based analyses of optical tweezers data ignore any possible motion blur \cite{BergSorenson2004}.
Comparison between Eq.~\ref{EQ-37} and \ref{TweezersMSD}
suggests that in the case of exponential time-dependence motion blur changes the interpretation of the fluctuation amplitude without changing the shape of the MSD or PSD.
However, this circumstance is special to the case of exponential relaxations and does not hold in general \cite{Jakeman1974}. (See also, for example, Sec.~\ref{BeadInPEO})


\section{MSD and Covariance matrix elements for fractional Brownian motion (1D)}
\label{AppendixE}
In the case of
fractional Brownian motion
(fBm), our starting point is
theoretical mean-square displacement from time $t_1$ to time $t_2$
\cite{Backlund2015}, namely
\begin{equation}
\left < [x(t_2) -  x(t_1)]^2 \right > = 2D | t_2 - t_1|^\alpha,
\label{TheoreticalMSD}
\end{equation}
where $\alpha$ is the exponent characterizing the fBm, and where we refer to $D$ as the diffusion coefficient, although its dimensions are m$^2$s$^{-\alpha}$.
Incorporating motion blur and static localization noise, the MSD becomes
\begin{widetext}
\begin{align}
M_n &= \frac{2D}{\Delta t_E^2} \int^{\Delta t_E}_0 dt \int_0^{\Delta t_E} ds [ | n \Delta t + t -s | ^\alpha  - |t-s|^\alpha ] +2\sigma^2
\nonumber \\
&= \frac{2D (\Delta t)^{2+\alpha}\left ( (n-\frac{\Delta t_E}{\Delta t})^{2+\alpha} -2 n^{2+\alpha}+ (n+\frac{\Delta t_E}{\Delta t})^{2+\alpha} \right )}
{(1+\alpha)(2+\alpha) \Delta t_E^2} -
\frac{
4 D \Delta t_E^{\alpha}
}
{
(1+\alpha)(2+\alpha)
} +2\sigma^2,
\label{BlurredMSDn}
\end{align}
which reproduces the result given in Ref.~\cite{Savin2005}.

%

The calculation for the fBm covariance terms can be performed as follows:
\begin{align}
\Sigma_n & = \int_{0}^{\Delta t_E} d s \int_{0}^{\Delta t_E} d t [x((n+1) \Delta t + t) - x(n \Delta t + t)][x(\Delta t + s) - x(s)]
\nonumber \\
&=  \int_{0}^{\Delta t_E} d s \int_{0}^{\Delta t_E} d t [x((n+1) \Delta t + t) x(\Delta t + s) - x((n + 1) \Delta t + t) x(s) - x(n \Delta t + t) x( \Delta t + s) + x(n \Delta t + t) x(s)]
\nonumber \\
&=  \int_{0}^{\Delta t_E} d s \int_{0}^{\Delta t_E} d t [2 x(n \Delta t + t) x(s) - x((n+1) \Delta t + t) x(s) - x(n \Delta t + t) x(\Delta t +s)]
\label{MLB01}
\end{align}

Condense to form $\left < [x(t_2) -  x(t_1)]^2 \right > = 2D | t_2 - t_1|^\alpha$, from Eq.~\ref{TheoreticalMSD}

\begin{equation}
\Sigma_n = \int_{0}^{\Delta t_E} d s \int_{0}^{\Delta t_E} d t [\frac{1}{2} [x((n+1) \Delta t + t) - x(s)]^2 + \frac{1}{2} [x((n-1) \Delta t + t) - x(s)]^2 - [x(n \Delta t + t) - x(s)]^2]
\label{MLB02}
\end{equation}

Plug in $2D | t_2 - t_1|^\alpha$:

\begin{equation}
\Sigma_n = \frac{D}{\Delta t_E^2} \int_{0}^{\Delta t_E} d s \int_{0}^{\Delta t_E} d t  [|n \Delta t + \Delta t + t- s|^\alpha + |n \Delta t - \Delta t + t - s|^\alpha - 2|n \Delta t +t - s|^\alpha]
\label{masterintegral}
\end{equation}

Now break Eq.~\ref{masterintegral} into 3 integrals, and calculate each one. Starting with the left-most integral:

\begin{align}
\int_{0}^{\Delta t_E} d s \int_{0}^{\Delta t_E} d t  |n \Delta t + \Delta t + t- s|^\alpha = \int_{0}^{\Delta t_E} d s \int_{0}^{\Delta t_E} d t  [n \Delta t + \Delta t + t- s]^\alpha
\end{align}

Absolute value can be ignored since $n \Delta t + \Delta t +t -s > 0$. The result is:
\begin{equation}
 \frac{-2(n \Delta t+ \Delta t)^{\alpha + 2}}{(\alpha +1)(\alpha + 2)} + \frac{(n \Delta t + \Delta t - \Delta t_E)^{\alpha+2}}{(\alpha +1)(\alpha + 2)} + \frac{(n \Delta t + \Delta t + \Delta t_E)^{\alpha + 2}}{(\alpha +1)(\alpha + 2)}
 \label{intpart1}
 \end{equation}

Next we have the middle integral:
\begin{equation}
\int_{0}^{\Delta t_E} d s \int_{0}^{\Delta t_E} d t |n \Delta t - \Delta t + t - s|^{\alpha}
\end{equation}

To evaluate this integral, we must take the cases when $n=1$ and $n>1$ into account. For $n=1$, the integrand becomes $|t-s|^{\alpha}$:

 \begin{equation}
\int_{0}^{\Delta t_E} d s \int_{0}^{\Delta t_E} d t |t-s|^{\alpha} = \frac{2 (\Delta t_E)^{\alpha + 2}}{(\alpha + 1)(\alpha+2)}
 \end{equation}

For $n>1$, $n \Delta t - \Delta t +t -s > 0$, so we can do the integral normally.
\begin{equation}
\int_{0}^{\Delta t_E} d s \int_{0}^{\Delta t_E} d t [n \Delta t - \Delta t + t - s]^{\alpha} = \frac{-2(n \Delta t - \Delta t)^{\alpha + 2}}{(\alpha +1)(\alpha + 2)} + \frac{(n \Delta t - \Delta t - \Delta t_E)^{\alpha + 2}}{(\alpha +1)(\alpha + 2)} + \frac{(n \Delta t - \Delta t + \Delta t_E)^{\alpha +2}}{(\alpha +1)(\alpha + 2)}
\end{equation}

Lastly, the third part of the integral can be done without the absolute value, since $n \Delta t > \Delta t_E$:
\begin{equation}
-2 \int_{0}^{\Delta t_E} d s \int_{0}^{\Delta t_E} d t [(n \Delta t + t - s)^{\alpha}] = \frac{4 (n \Delta t)^{\alpha +2}}{(\alpha +1)(\alpha + 2)} - \frac{2(n\Delta t-\Delta t_E)^{\alpha +2}}{(\alpha +1)(\alpha + 2)} - \frac{2(n\Delta t+\Delta t_E)^{\alpha+2}}{(\alpha +1)(\alpha + 2)}
\end{equation}

Thus, the covariance terms are:

\begin{equation}
\Sigma_0 =  \frac{2D (\Delta t)^{2+\alpha}\left ( (1-\frac{\Delta t_E}{\Delta t})^{2+\alpha} -2+ (1+\frac{\Delta t_E}{\Delta t})^{2+\alpha} \right )}
{(1+\alpha)(2+\alpha) \Delta t_E^2} -
\frac{
4 D \Delta t_E^{\alpha}
}
{
(1+\alpha)(2+\alpha)
}
\label{eq:S0_fBM}
\end{equation}

\begin{equation}
\begin{split}
\Sigma_1 = \frac{D}{\Delta t_E^2} [ \frac{-2(2\Delta t)^{\alpha + 2}}{(\alpha +1)(\alpha + 2)} + \frac{(2 \Delta t - \Delta t_E)^{\alpha+2}}{(\alpha +1)(\alpha + 2)} + \frac{(2 \Delta t + \Delta t_E)^{\alpha + 2}}{(\alpha +1)(\alpha + 2)} +\frac{2 (\Delta t_E)^{\alpha + 2}}{(\alpha + 1)(\alpha+2)}
\\ +\frac{4 (\Delta t)^{\alpha +2}}{(\alpha +1)(\alpha + 2)} - \frac{2(\Delta t-\Delta t_E)^{\alpha +2}}{(\alpha +1)(\alpha + 2)} - \frac{2(\Delta t+\Delta t_E)^{\alpha +2}}{(\alpha +1)(\alpha + 2)} ]
\end{split}
\label{eq:S1_fBM}
\end{equation}

\begin{equation}
\begin{split}
\Sigma_n =  \frac{D}{\Delta t_E^2}[\frac{-2(n \Delta t+ \Delta t)^{\alpha + 2}}{(\alpha +1)(\alpha + 2)} + \frac{(n \Delta t + \Delta t - \Delta t_E)^{\alpha+2}}{(\alpha +1)(\alpha + 2)} + \frac{(n \Delta t + \Delta t + \Delta t_E)^{\alpha + 2}}{(\alpha +1)(\alpha + 2)}
\\ -\frac{2(n \Delta t - \Delta t)^{\alpha + 2}}{(\alpha +1)(\alpha + 2)} + \frac{(n \Delta t - \Delta t - \Delta t_E)^{\alpha + 2}}{(\alpha +1)(\alpha + 2)} + \frac{(n \Delta t - \Delta t + \Delta t_E)^{\alpha +2}}{(\alpha +1)(\alpha + 2)}
\\ +\frac{4 (n \Delta t)^{\alpha +2}}{(\alpha +1)(\alpha + 2)} - \frac{2(n\Delta t-\Delta t_E)^{\alpha +2}}{(\alpha +1)(\alpha + 2)} - \frac{2(n\Delta t+\Delta t_E)^{\alpha+2}}{(\alpha +1)(\alpha + 2)}]
\end{split}
\label{eq:Sn_fBM}
\end{equation}
For $n=2,3,4,...$
\end{widetext}

In the limit that $\Delta t_E = 0$,
\begin{equation}
\Sigma_0 =  2D(\Delta t)^\alpha,
 \label{eq:S0_fBM_tE0}
\end{equation}

\begin{equation}
\Sigma_1 =  D[(2\Delta t)^\alpha-2(\Delta t)^\alpha],
 \label{eq:S1_fBM_tE0}
\end{equation}

\begin{equation}
\Sigma_n =  D[(n\Delta t+\Delta t)^\alpha+(n\Delta t-\Delta t)^\alpha-2(n\Delta t)^\alpha],
 \label{eq:Sn_fBM_tE0}
\end{equation}
For $n=2,3,4,...$

In the limit that $\alpha = 1$,
\begin{equation}
\Sigma_0 =  2D\Delta t-\frac{2D\Delta t_E}{3},
 \label{eq:S0_fBM_a1}
\end{equation}

\begin{equation}
\Sigma_1 =  \frac{D\Delta t_E}{3},
 \label{eq:S1_fBM_a1}
\end{equation}

\begin{equation}
\Sigma_n =  0,
 \label{eq:Sn_fBM_a1}
\end{equation}
for $n \geq 2 $.

\section{Additional figures}

\subsubsection{Simulated fractional Brownian motion with uniform localization noise}
\label{fBmHomogen}

\begin{figure} 
	\includegraphics[scale=0.450]{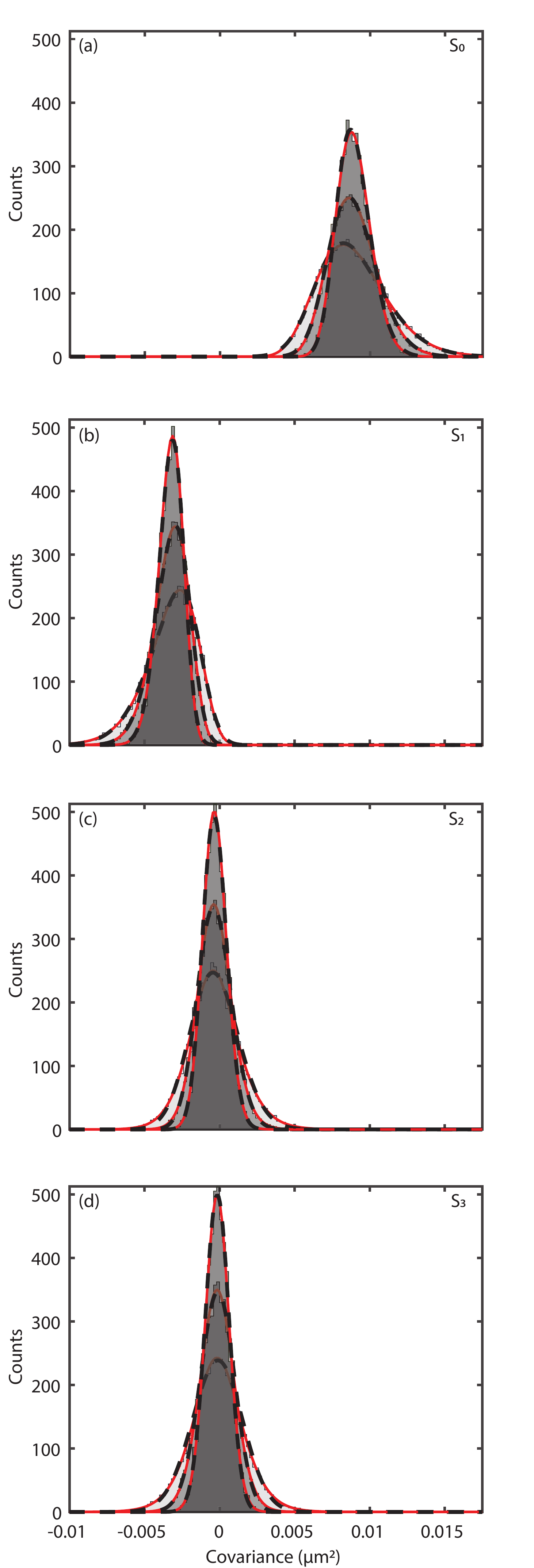}
	\caption{\label{fig:fig1_fbmUN} Covariance distributions for simulated fBm tracks with uniform localization noise, $D = 0.0055~\mu$m$^2$s$^{-1}$, $\sigma^2 = 0.0032~\mu$m$^2$, and $\Delta t = \Delta t_E = 0.058s$. Covariances ${S}_0$ (a), ${S}_1$ (b), ${S}_2$ (c), and ${S}_3$ (d) for particle tracks of 19 (light gray), 39 (medium gray), and 79 (dark gray) steps are represented as histograms. Red lines correspond to the theoretical distributions. Black dashed lines correspond to the best fit of a skew normal distribution to the simulated distributions. With increasing number of time steps, the distribution narrows.  }
\end{figure}

\begin{figure}
	\includegraphics[scale=0.45]{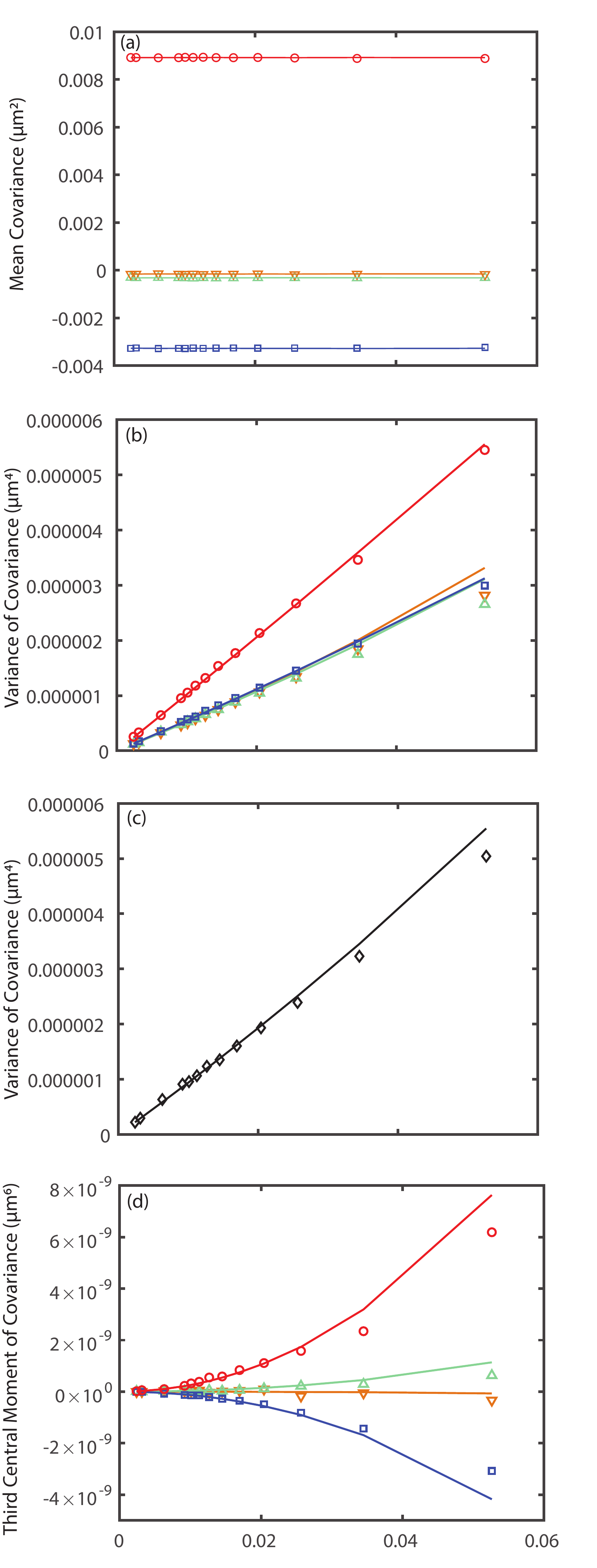}
	\caption{\label{fig:fig2_fBmUN} Dependence of the covariance matrix estimates on the track length for simulated fBm tracks with uniform localization noise. (a) The mean of the covariances $S_0$ (red circle),  $S_1$ (blue square),  $S_2$ (green triangle), and  $S_3$ (orange upside-down triangle) vs. inverse track length. Theory (straight lines) is calculated by Eq.~\ref{mean}. (b) Variance of the covariance vs. inverse track size of  ${S}_0$ (red circle),  ${S}_1$ (blue square),  ${S}_2$ (green triangle), and  ${S}_3$ (orange upside-down triangle). The theory is calculated by Eq.~\ref{Variance2D}. (c) Variance of the covariance vs. inverse bin size of ${S}_0 + 2{S}_1$ (diamond). Theory (solid line) is calculated using $\frac{{\rm Tr}({\bf \Sigma} ({\bf C}_0+2{\bf C}_1))^2}{4}$ (d) The third central moment of the covariance ${\Sigma}_0$ (circle),  ${\Sigma}_1$ (square),  ${\Sigma}_2$ (triangle), and  ${\Sigma}_3$ (upside-down triangle) vs. inverse track length. Theory (lines) is calculated by Eq.~\ref{Skewness2D}. } 
\end{figure}

\nocite{*}

\bibliography{apssamp_MLB+SM042220}

\end{document}